\documentclass[aps,prb,10pt,twocolumn,amsfonts,superscriptaddress,showpacs]{revtex4-1}

\usepackage{color}
\usepackage{bm}
\usepackage{bbold}
\usepackage{amsmath}
\usepackage{amssymb}
\usepackage{epsfig}
\usepackage{verbatim}
\usepackage[T1]{fontenc}
\usepackage{array}

\newcommand{\be}{\begin{equation}}
\newcommand{\ee}{\end{equation}}
\newcommand{\bea}{\begin{eqnarray}}
\newcommand{\eea}{\end{eqnarray}}
\newcommand{\la}{\langle}
\newcommand{\ra}{\rangle}
\newcommand{\dg}{^\dagger}
\newcommand{\p}{\partial}
\newcommand{\rd}{{\rm d}}
\def\nn{\nonumber\\}
\def\fr#1{(\ref{#1})}
\def\ocite#1{[\onlinecite{#1}]}

\usepackage[colorlinks=true,bookmarks=false,citecolor=blue,linkcolor=blue,hyperfootnotes=true]{hyperref}

\begin{document}

\title{Motion of a distinguishable impurity in the Bose gas: Arrested  expansion without a lattice and impurity snaking}

\author{Neil J. Robinson}
\email{nrobinson@bnl.gov}
\affiliation{CMPMS Dept., Brookhaven National Laboratory, Upton, NY 11973-5000, USA}
\affiliation{The Rudolf Peierls Centre for Theoretical Physics, University of Oxford, Oxford, OX1 3NP, United Kingdom}

\author{Jean-S\'ebastien Caux}
\affiliation{Institute for Theoretical Physics, University of Amsterdam, Science Park 904,
Postbus 94485, 1090 GL Amsterdam, The Netherlands}

\author{Robert M. Konik}
\affiliation{CMPMS Dept., Brookhaven National Laboratory, Upton, NY 11973-5000, USA}

\date{\today}

\begin{abstract}
We consider the real time dynamics of an initially localized distinguishable impurity injected into the 
ground state of the Lieb-Liniger model. Focusing on the case where integrability is preserved, we
numerically compute the time evolution of the impurity density operator in regimes far from analytically
tractable limits. We find that the injected impurity undergoes a stuttering motion as it moves and expands.
For an initially stationary impurity, the interaction-driven formation of a quasibound state with a hole 
in the background gas leads to arrested expansion -- a period of quasistationary behavior. When the impurity
is injected with a finite center of mass momentum, the impurity moves through the background gas in a snaking 
manner, arising from a quantum Newton's cradle-like scenario where momentum is exchanged back-and-forth 
between the impurity and the background gas. 
\end{abstract}


\maketitle

\textit{Introduction}.---
With recent experimental advances in the field of cold atomic gases, the physics of the one-dimensional Bose gas is receiving an increasing amount of attention.\cite{ParedesNature04,KinoshitaScience04,WeissNature06,CauxPRA06,HofferbethNature07,PalzerPRL09,WillPRA11,CazalillaRMP11,CataniPRA12,JohnsonEPL12,TrotzkyNatPhys12,MasselNJP13,VolsnievArxiv15,CartariusArxiv15,LangenScience15} These systems, in which one has unprecedented isolation from the environment and fine control of interparticle interactions, are excellent tools for examining novel phenomena arising from strong correlations. 

One such phenomenon which has piqued both theoretical\cite{Heidrich-MeisnerPRA09,CazalillaRMP11,MuthPRA12,JreissatyPRA13,RonzheimerPRL13,VidmarPRB13,CartariusArxiv15,VidmarArxiv15} and experimental\cite{JendrzejewskiNatPhys12,RonzheimerPRL13} curiosity is the expansion dynamics of a gas of cold atoms. A number of surprising and interesting effects have been observed, of which arrested expansion (or self trapping)\cite{Heidrich-MeisnerPRA09,MuthPRA12,JreissatyPRA13} is of particular relevance to this work. A gas (bosons\cite{RonzheimerPRL13} or fermions\cite{SchneiderNatPhys12,SchmidtPRL13}) is released from a confining potential and allowed to expand on the lattice. Under this time evolution `bimodal' expansion is observed: the sparse outer regions of the cloud rapidly expand whilst the dense central region spreads only very slowly. This can be partially understood by considering the limit of strong interactions: doubly occupied sites are high energy configurations which, thanks to the lattice imposing a finite bandwidth and energy conservation, cannot release their energy to the rest of the system and decay.\cite{JreissatyPRA13}

With these recent experimental advances\cite{LewensteinAdvPhys07} has also come the ability to examine systems in which there is a large imbalance between two species;\cite{KleinNJP07,PalzerPRL09,SchirotzekPRL09,BrudererPRA10,WillPRA11,CataniPRA12,MassignanRepProgPhys14} a natural starting point for the study of impurity physics. This gives insight in to a diverse range of problems,\cite{MassignanRepProgPhys14} from the physics of polarons\cite{FeynmanPR55,SchirotzekPRL09} to the x-ray edge singularity\cite{MahanPR67,NozieresPR69} and the orthogonality catastrophe.\cite{AndersonPRL67} The physics of impurities also plays an important role in the calculation of edge exponents in dynamical correlation functions\cite{ImambekovRMP12} and in understanding the nonequilibrium dynamics following a local quantum quench.\cite{ZvonarevPRL07,GanahlPRL12,VolsnievArxiv15}

The experimental study of the out-of-equilibrium dynamics of a single impurity in the one-dimensional Bose gas has revealed rather rich physics: from how an impurity spreads when accelerated through a Tonks-Girardeau gas,\cite{PalzerPRL09} to how interactions effect oscillations in the size of a trapped out-of-equilibrium impurity.\cite{CataniPRA12} Numerous theoretical investigations have addressed the Tonks-Girardeau regime: from a static point impurity\cite{GooldNJP10,GooldPRA11} to a completely delocalized (e.g., plane wave) impurity.\cite{MathyNatPhys12,KnapPRL14,BurovskiPRA14,GamayunPRA14,GamayunPRE14,LychkovskiyPRA2014} Away from the Tonks-Girardeau limit, theoretical study of the continuum problem is challenging and results have focused on lattice models, such as the Bose-Hubbard model.\cite{KleinePRA08,CazalillaRMP11}

In this letter we consider the out-of-equilibrium dynamics of an initially localized impurity in the Lieb-Liniger model. Using a combination of exact analytical results and numerical computations, we show that an impurity injected into the ground state of the Lieb-Liniger model undergoes a stuttering sequence of rapid movement/expansion followed by arrested expansion. For an initially stationary impurity, this is caused by the interaction-driven out-of-equilibrium formation of a quasibound state of the impurity with a hole in the background gas. This quasibound state is robust under time-evolution for long periods of time. For an impurity with a finite initial center of mass (COM) momentum, the stuttering sequence results in the impurity ``snaking'' through the background gas; the impurity exchanges momentum back-and-forth with the background gas through a quantum Newton's cradle-like mechanism.\cite{WeissNature06} The results we present are relevant to experiments (see, e.g. Ref.~\ocite{PalzerPRL09}) and should be observed under reasonable conditions.

\textit{The two-component Lieb-Liniger model}.---
We consider two species of delta function interacting bosons confined to a ring of length $L$. The Hamiltonian of the two-component Lieb-Liniger model (TCLLM) is given by 
\bea
H &=& \int_0^L \rd x  \sum_{j=1,2} \frac{\hbar^2}{2m} \p_x\Psi_j\dg (x)\p_x\Psi^{\phantom\dagger}_j(x)\nn 
&& +  \int_0^L \rd x \sum_{j,l = 1,2} c \Psi\dg_j(x)\Psi\dg_l(x)\Psi^{\phantom\dagger}_l(x)\Psi^{\phantom\dagger}_j(x),\label{Eq:H_LL}
\eea
where herein we set $\hbar = 2m = 1$, $c$ is the interaction parameter, and the boson operators obey the canonical commutation relations $[\Psi_j(x), \Psi\dg_l(y) ] = \delta_{j,l}\delta(x-y)$ with $j,l=1,2$ denoting the species. As in the case of the one-component Lieb-Liniger model,\cite{LiebPR63,LiebPR63a,KorepinBook} the generalization to multiple particle species remains integrable provided all species interact identically.\cite{YangPRL67,SutherlandPRL68}  

The TCLLM can be solved by the Bethe Ansatz,\cite{YangPRL67,SutherlandPRL68} giving access to some of its basic physical properties (see, e.g., \ocite{CauxPRA09,KlauserPRA11,PozsgayJPhysA12} and references therein). An $N$-particle eigenstate containing $N_1$ particles of species 1 is characterized by a set of $N$ momenta $\{q\}_N = \{q_1,\ldots,q_N\}$ and a set of $N_1$ species rapidities $\{\lambda\}_{N_1} = \{\lambda_1,\ldots,\lambda_{N_1}\}$.  These momenta and rapidities satisfy the nested Bethe ansatz equations
\bea
&&e^{iq_jL} = -\prod_{l=1}^N \frac{q_j-q_l+ic}{q_j-q_l-ic} \prod_{m=1}^{N_1} \frac{q_j - \lambda_m - \frac{ic}{2}}{q_j - \lambda_m + \frac{ic}{2}}\ ,\label{Eq:BA1}\\
&&\prod_{l=1}^N \frac{\lambda_k - q_l - \frac{ic}{2}}{\lambda_k - q_l + \frac{ic}{2}} = -\prod_{l=1}^{N_1} 
\frac{\lambda_k - \lambda_l - ic}{\lambda_{k}-\lambda_l + ic}\ ,\label{Eq:BA2}
\eea
where $j = 1,\ldots, N$ and $k=1,\ldots,N_1$. The eigenstate $|\{q\}_N;\{\lambda\}_{N_1}\ra$ has energy $E_{q} = \sum_j q_j^2$ and  momentum $K_{q} = \sum_j q_j$.

\textit{The initial state}.---
We study the dynamics of an impurity starting from the state 
\be
|\Psi(Q)\ra = \frac{1}{{\cal N}} \int_0^L \rd x\ e^{iQx}e^{-\frac12\left(\frac{x-x_0}{a_0} \right)^2} \Psi\dg_1(x) |\Omega\ra,
\label{Eq:Initial_State}
\ee
where $|\Omega\ra$ is the $N_2 = N-1$ particle ground state of the one-component Lieb-Liniger model, $Q$ is the COM momentum of the impurity and ${\cal N}$ normalizes the state. The study of such a state is partially motivated by the experiments performed in Refs.~\onlinecite{PalzerPRL09,CataniPRA12}, which study the dynamics of an impurity in a background gas.

The initially localized impurity of Ref.~\onlinecite{PalzerPRL09} is prepared by illuminating a trapped one-component Bose gas with a radio-frequency pulse; this causes transitions between the $|F,m_F\ra = |1,-1\ra$ hyperfine state of the trapped gas and the $|1,0\ra$ state (the impurity). Due to the magnetic trap, transitions occur only within a spatially localized region, the thinness of which is Fourier-limited by the pulse duration. The resulting impurity contains up to three particles and is accelerated through the gas by gravity, as the $|1,0\ra$ state does not experience the magnetic trap.

On the other hand, the impurity in Ref.~\onlinecite{CataniPRA12} is prepared by first tuning the interspecies interaction to zero and then using a species-dependent trap and light blade to shape the impurity. Following this preparation, the interspecies interaction is turned on and the impurity released from the trap/light blade and its expansion studied. 

To distill the intrinsic dynamics of the impurity, our scenario varies slightly from experiments~\cite{PalzerPRL09,CataniPRA12}: we study an impurity injected into a constant density background gas in the absence of an external potential (such as a magnetic trap and gravity). Similar approximations have been applied in the well-studied yrast states.\cite{KaminishiPRA11,SatoPRL12,SatoArxiv12,KaminishiArxiv13,KaminishiArxiv14}

\textit{Time evolution protocol.---}
Our aim is to compute the impurity density profile when the initial state~\fr{Eq:Initial_State} is time-evolved according to the Hamiltonian~\fr{Eq:H_LL} $\rho_1(x,t) = \la \Psi(Q)|e^{i Ht} \Psi\dg_1(x)\Psi^{\phantom\dagger}_1(x) e^{-i Ht} |\Psi(Q)\ra.$ This is a nontrivial problem as the initial state~\fr{Eq:Initial_State} \textit{is not an eigenstate} of the Hamiltonian. We use the integrability of the TCLLM to numerically evaluate the density profile using recently derived results for matrix elements of local operators.\cite{PozsgayJPhysA12} Due to a dearth of results for matrix elements in the TCLLM, we are restricted to studying the density of the impurity and we cannot examine the background gas.\footnote{After the completion of this work, new results for matrix elements in the TCLLM were obtained.\cite{PakuliakArxiv15} We hope to incorporate these results in future works and note that these new expressions remove our limitation of only studying the impurity density.}

The essential idea is the following: we insert complete sets of eigenstates between each time evolution operator and the initial state in $\rho_1(x,t)$. By orthogonality, we sum over the Bethe states with $N_1 = 1$ and $N_1 + N_2 = N$. The momenta and rapidities characterizing these states satisfy the nested Bethe ansatz equations (\ref{Eq:BA1},\ref{Eq:BA2}). The density profile of the impurity will then be given by
\bea	
\rho_1(x,t) &=& \sum_{\substack{\{k\};\mu \\ \{p\};\lambda } }  
\la\Psi(Q)|\{p\};\lambda\ra \la \{p\};\lambda| \Psi\dg_1(0)\Psi^{\phantom\dagger}_1(0) |\{k\};\mu\ra\nn 
&&\qquad  \times  \la\{k\};\mu|\Psi(Q)\ra e^{i (E_{p}-E_{k})t} e^{i(K_{p}-K_{k})x}  ,
\label{Eq:FFExpansion}
\eea
where $\{q\} \equiv \{q\}_N$. The overlap of the initial state with a Bethe state can be expressed as ${\cal N} \la \{k\};\mu|\Psi(Q)\ra = \int_0^L \rd x\ e^{i(Q + K_{k}-K_\Omega)x} e^{-\frac12\left(\frac{x-x_0}{a_0}\right)^2}\la \{k\};\mu| \Psi\dg_1(0)|\Omega\ra$, where $K_\Omega$ is the momentum of the ground state $|\Omega\ra$. So, in order to compute~\fr{Eq:FFExpansion} we require two ingredients: the matrix elements of the creation operator $\Psi\dg_1(0)$ and the density operator $\Psi\dg_1(0)\Psi^{\phantom\dagger}_1(0)$ on the Bethe states. These matrix elements have been derived from the algebraic Bethe ansatz.\cite{PozsgayJPhysA12,PakuliakArxiv15} Required results are summarized in the Supplemental Materials.\cite{SuppMat}

Readers interested in our scheme for numerically evaluating the expansion~\fr{Eq:FFExpansion} can consult Ref.~\ocite{CauxJMathPhys09}. An important point to note is that the expansion~\fr{Eq:FFExpansion} contains an \textit{infinite} number of terms.  We truncate the Hilbert space by selecting the Bethe states which have the largest overlaps with the initial state~\fr{Eq:Initial_State}. To quantify the truncation error, we compute the saturation of the sum rule
\be
\sum_{\{k\}_N;\mu} \Big|\la\Psi(Q)|\{k\}_N;\mu\ra\Big|^2 = 1,
\label{Eq:SumRule}
\ee	
and we present numerical values for this with our results. We are limited to small numbers of particles $N \lesssim 10$ and we have to keep $\sim10^4 - 10^5$ states to saturate the sum rule to 2 decimal places. 

\textit{The noninteracting limit}.---
In the noninteracting limit, the time evolution of the initial state~\fr{Eq:Initial_State} is a single particle problem. The time-dependent density profile can be calculated exactly (we take $L\to\infty$): $\rho_1(x,t)_{c=0} = \frac{a_0}{\sqrt{\pi}} \exp(-\frac{a_0^2 (x+2Qt)^2}{a_0^4 + t^2})/{\sqrt{ a_0^4 + t^2}}$. The noninteracting density profile remains Gaussian at all times, with a time-dependent width and amplitude. 

\begin{figure}
\includegraphics[trim=140 110 40 180,clip,width=0.4\textwidth]{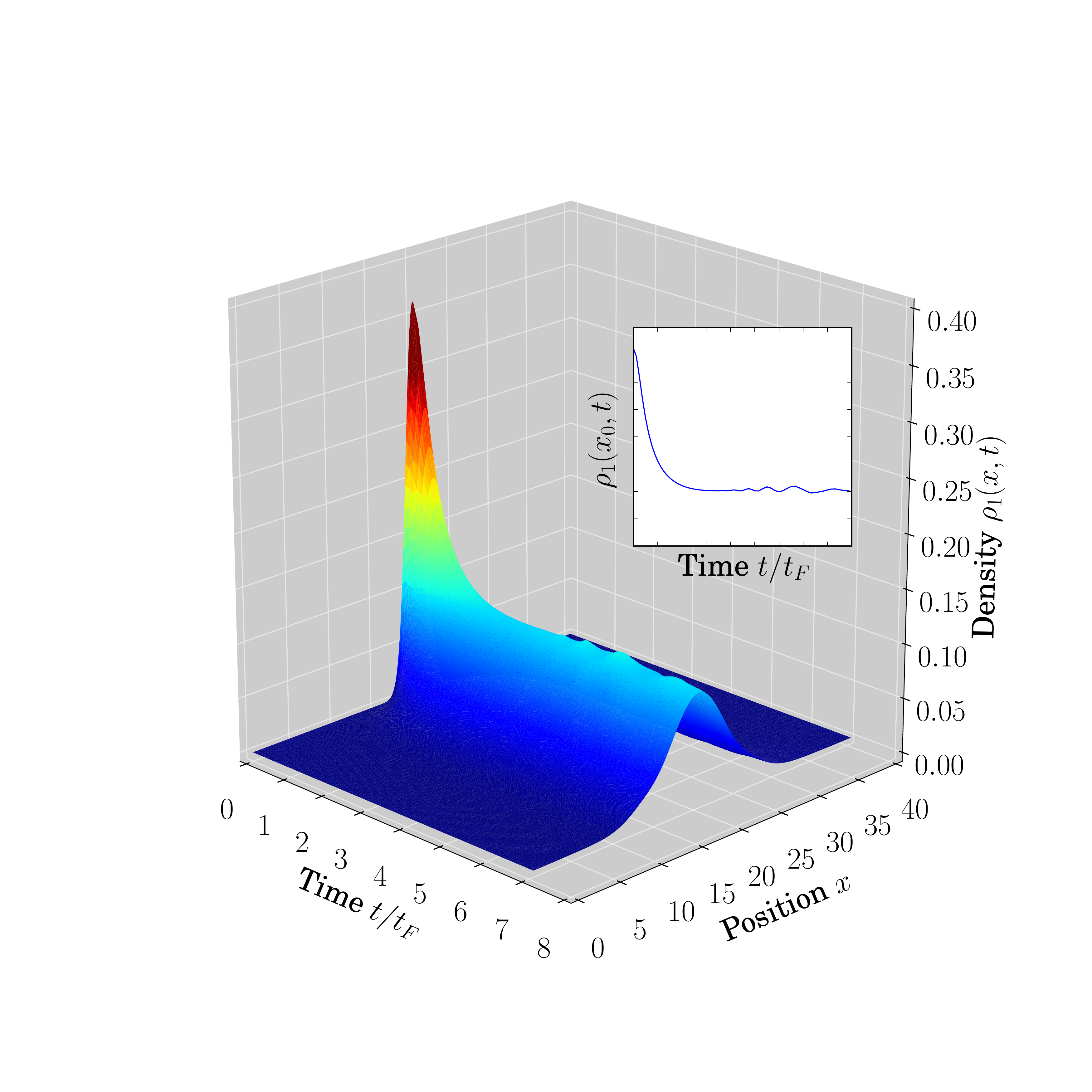} 
\caption{Time evolution of the impurity density of the initial state~\fr{Eq:Initial_State} with $Q=0$, $x_0=L/2$ and $a_0^2 = 1.125$
on the $L=40$ ring for a system of $N=8$ particles with interaction parameter $c=10$. The Hilbert space is truncated
to 25150 states, leading to the sum rule~\fr{Eq:SumRule}$\ =0.9858$. (Inset) Time-evolution of the maximum of the density $\rho_1(x_0,t)$.
Constant time cuts can be found in the Supplemental Materials.\cite{SuppMat}}
\label{Fig:Gaussian_Q0}
\end{figure}

\textit{Arrested expansion: $Q=0$.---} 
In Fig.~\ref{Fig:Gaussian_Q0} we present results for the time evolution of the impurity density profile~\fr{Eq:FFExpansion} for $N=8$ bosons on the circumference $L=40$ ring starting from the initial state~\fr{Eq:Initial_State} with  $x_0 =L/2$, $a_0^2 =1.125$ and interaction parameter $c=10$. We measure time in units of $t_F = 1/E_F$ where $E_F = (\pi N/L)^2$ is the Fermi energy in the $c\to\infty$ limit. The Hilbert space is truncated to 25150 states, leading to the sum rule saturation $0.9858$ (i.e., to $1.4\%$). Upon time evolution the wave packet spreads, maintaining its Gaussian shape as in the noninteracting case. However, at time $t\sim2t_F$ the wave packet stops spreading and only undergoes small amplitude breathing oscillations. This arrested expansion is an example of \textit{prethermalization}.\cite{MoeckelPRL08,RoschPRL08,MoeckelAnnPhys09,KollarPRB11,MarcuzziPRL13,EsslerPRB14,MenegozJStatMech15,BertiniArxiv15} The system relaxes in a two-step process, first approaching a quasi-stationary non-equilibrium state (the arrested expansion) before subsequent equilibration.  Two-step relaxation has been observed in the one-dimensional Bose gas following a global quantum quench.\cite{GringScience12,LangenEPJST13,AduSmithNJP13}

We can qualitatively reproduce aspects of this behavior with a mean field (MF) decoupling of the interaction term
\be
\Psi\dg_j(x)\Psi\dg_l(x)\Psi^{\phantom\dagger}_l(x)\Psi^{\phantom\dagger}_j(x) \approx \rho_j(x,t) \Psi\dg_l(x)\Psi^{\phantom\dagger}_l(x) + j\leftrightarrow l. 
\label{Eq:Mean_Field}
\ee
At a MF level the impurity profile is a time-dependent repulsive one-body potential for the background gas. The region under the impurity then excludes particles in the background gas, resulting in the formation of a `hole'. This hole in the background gas acts as a confining (attractive) one body potential for the impurity in MF and the two form a quasibound particle-hole pair,\footnote{See~[\onlinecite{SuppMat}] for a mean-field analysis in a discretized model which shows this picture captures the correct physics.} much like an exciton in the electron gas (see, e.g., Ref.~\ocite{KnoxExcitons}). This is different to the self-trapping scenario on the lattice: there the `doublons' are stable as the large interaction energy cannot be converted into kinetic energy due to particle number conservation and the finite bandwidth. In the continuum, dynamical arrest is driven by the formation of the impurity-hole quasibound state and is not observed for an indistinguishable impurity.\cite{SuppMat} 

At later times ($t\gtrsim 7t_F$), the impurity eventually broadens. This broadening occurs in a sequence of expansion/arrested expansion steps, whilst the impurity undergoes small amplitude breathing oscillations.\cite{SuppMat} The slow decay of the density at later times may be related to the subdiffusive equilibrium behavior reported in Ref.~\ocite{ZvonarevPRL07}. However, finite-size effects and our choice of observable obscure the characteristic logarithmic decay of subdiffusion.

\begin{figure}
\includegraphics[trim=140 75 40 125,clip,width=0.39\textwidth]{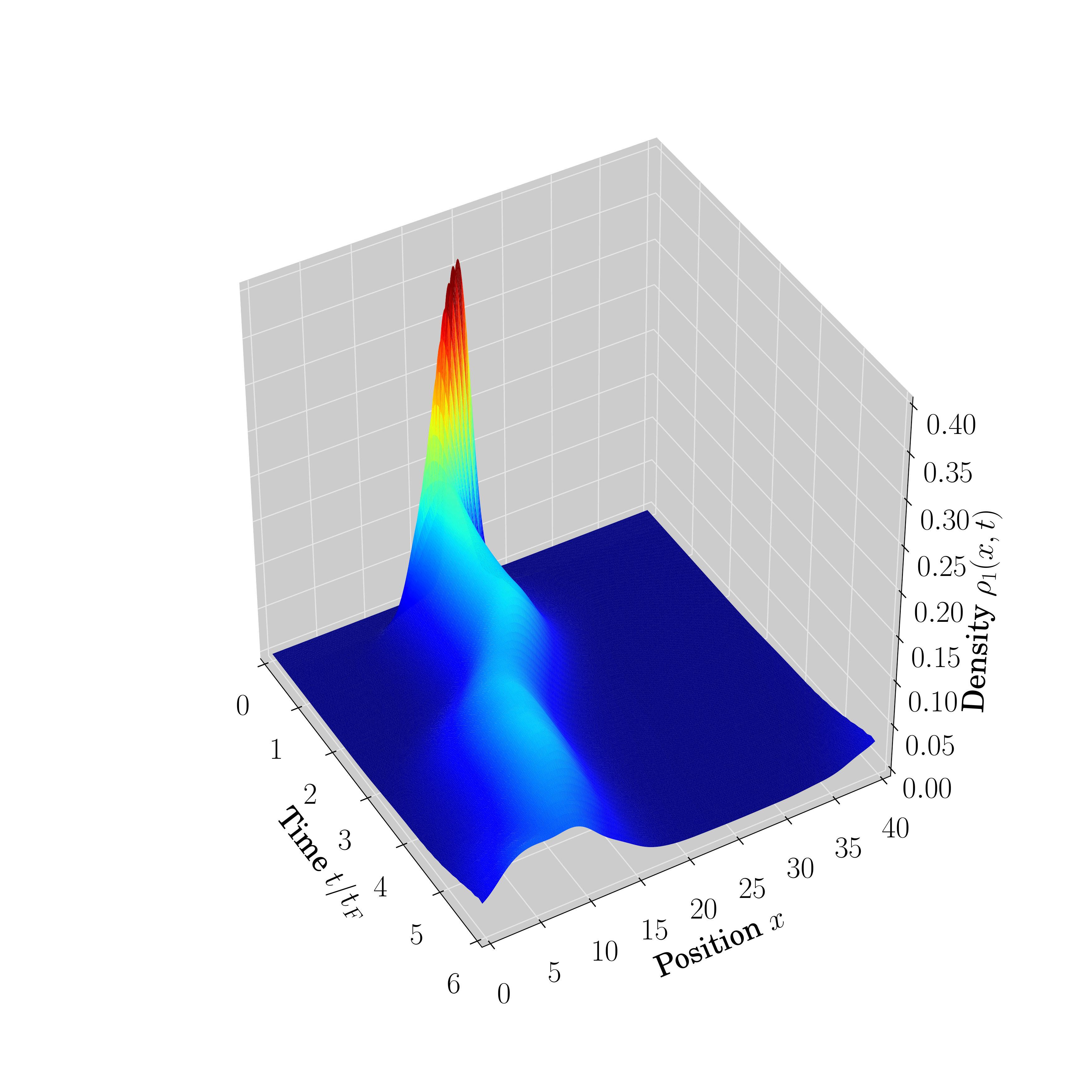}\\		
\caption{The dynamics of the impurity density~\fr{Eq:FFExpansion} from the initial state~\fr{Eq:Initial_State} with $a_0^2 = 1.125$ and $Q=40\pi/L$. We use 21,507 states to study a system of $N=8$ bosons on the length $L=40$ ring with interaction parameter $c=10$, resulting in the sum rule~\fr{Eq:SumRule}$\ = 0.981$. Plots of constant time cuts are presented in the Supplemental Materials. }
\label{Fig:Gaussian_Q}
\end{figure}

\begin{figure}[h]
\includegraphics[width=0.44\textwidth]{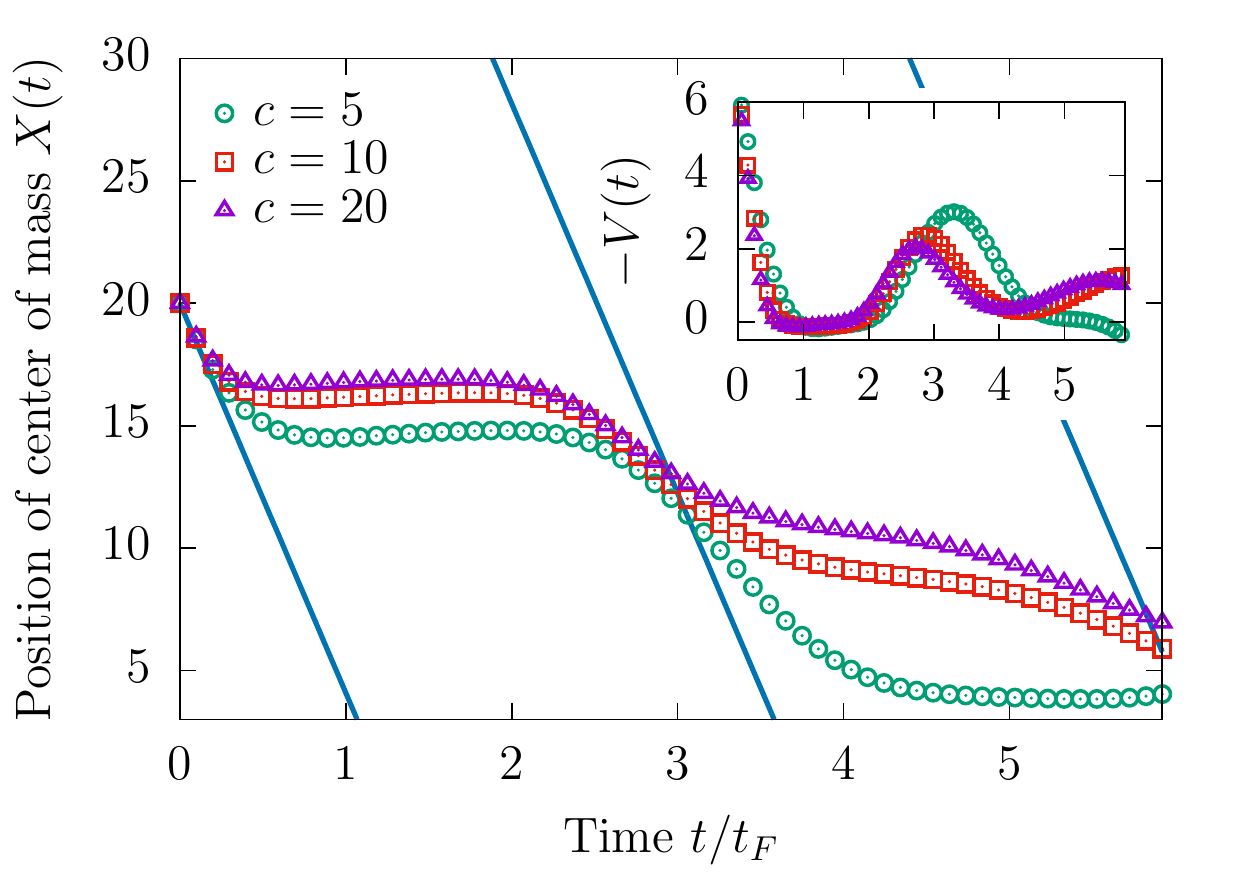}
\vspace{-0.25cm}
\caption{The time evolution of the center of mass $X(t)$~\fr{Eq:CoM}  for: (points) the initial state~\fr{Eq:Initial_State}
with $a_0^2 = 1.125$, $Q = \pi$ for $N=8$ particles on the length $L=40$ ring with interaction parameter $c=5,10,20$;
(line) a noninteracting point particle with mass $m=1/2$ and velocity $Q/m$. Inset: velocity of the center of mass 
$V(t) = \Delta X(t)/ \Delta t$ ($cf.$ Ref.~\onlinecite{MathyNatPhys12}).} 
\label{Fig:Center_of_mass}
\end{figure}

\textit{The snaking impurity: $Q\neq0$.---} 
Finally, we consider the time evolution of the initial state~\fr{Eq:Initial_State} with nonzero COM  momentum $Q$. Our prescription for computing the time evolution is identical to the $Q=0$ case; in Fig.~\ref{Fig:Gaussian_Q} we present results for the impurity density profile for the same set of parameters as in Fig.~\ref{Fig:Gaussian_Q0} with $Q = \pi$. We see rather surprising behavior: the impurity moves in a snaking manner, repeatedly moving and expanding before becoming approximately stationary with arrested expansion. To quantify the nonuniform motion of the impurity further, we define the COM coordinate $X(t)$ as
\be
X(t) = \frac{L}{2\pi} \arctan\left[\frac{\int_0^{2\pi} \rd \theta \sin\theta \rho_1(\theta,t)}{\int_0^{2\pi} \rd \theta \cos \theta \rho_1(\theta,t)} \right],
\label{Eq:CoM}
\ee
where $\theta = 2\pi x/L$. We plot the COM coordinate in Fig.~\ref{Fig:Center_of_mass} for a number of interaction strengths; $X(t)$ shows regions of rapid movement, followed by (approximately) stationary plateaux. Only at $t\lesssim t_F/3$ does the COM move as in the noninteracting case: $X(t)_{c=0} = X(0) - 2 Q t$. The sharpness of the plateaux and transient regions are governed by the interplay between the delocalization of the impurity and its interactions with the background gas.\footnote{In [\onlinecite{SuppMat}] we show that for weak interactions the impurity has almost completely delocalized by the second plateau, leading to a flat and stationary COM. On the other hand, with strong interactions the spreading of the impurity is hindered and on the second plateau the impurity is still well localized. Slight spreading of the (almost) stationary impurity leads to the drifting of the COM observed in Fig.~\ref{Fig:Center_of_mass}.}

We have the following picture for the behavior shown in Fig.~\ref{Fig:Gaussian_Q}: (i) The impurity moves to the left, scattering particles in the background gas and creating excitations with finite momentum. (ii) The impurity continues to scatter with the background until it imparts most (or all) of its COM momentum. (iii) The excitations in the background gas propagate around the ring and then collide once more with the impurity. (iv) The impurity gains COM momentum and the process repeats. In support of this picture is the behavior of the COM position plateau with system size $L$: the time for leaving the plateau $\tau_p$ is (approximately) linearly dependent on $L$. $\tau_p$ is also related to the initial momentum $Q$ of the impurity; for large $Q$, $\tau_p \sim 1/Q$ (an excitation with momentum $Q$ has velocity $\sim Q/m$). This $Q$-dependence reflects the momentum imparted by the impurity to excitations in the background gas, which then propagate around the ring.\footnote{See [\onlinecite{SuppMat}] for data showing the  length of time for which the COM coordinate is approximately stationary is determined by the total system size $L$ and the momentum $Q$ of the impurity.}  This process can be thought of in terms of a quantum Newton's cradle\cite{WeissNature06} on a ring, with the impurity exchanging momentum back-and-forth with the background gas, resulting in the snaking motion shown in Fig.~\ref{Fig:Gaussian_Q}.  In the Supplemental Materials we show that this behavior is not realized on the lattice when we perform the MF decoupling~\fr{Eq:Mean_Field} for the same set of parameters that capture some aspects of the $Q=0$ behavior.

It is interesting to consider removing periodic boundary conditions: excitations produced by injecting the impurity will propagate towards the boundary and subsequently reflect, returning to once again scatter the impurity. This reflection of the excitations means that we expect the COM to snake back-and-forth about $x_0$ rather than around the ring. In the presence of a harmonic trap, the COM will travel in a snaking motion due to both the trap and collisions with the background excitations.

A  question that has recently attracted attention is whether an injected impurity has finite momentum in the $t\to\infty$ limit (see, e.g., Refs.~[\onlinecite{MathyNatPhys12,KnapPRL14,BurovskiPRA14,GamayunPRE14,GamayunPRA14,LychkovskiyPRA2014}]). To address this, we compute the momentum of the impurity in the diagonal ensemble (DE) \footnote{For further details on our computations in the diagonal ensemble, see~[\onlinecite{SuppMat}].}  $ K_{DE} = \sum_{\{k\};\mu} \la \Psi(Q)| \{k\};\mu \ra\la\{k\};\mu|\Psi(Q)\ra \times \sum_p   \la\{k\};\mu| p \Psi\dg_{1,p}\Psi^{\phantom\dagger}_{1,p}| \{k\};\mu\ra,$ where $\Psi_{1,p} = 1/L \int \rd x e^{-ipx} \Psi_1(x)$. Doing so, we find $K_{DE} \approx -0.022$ (for $N=4$ particles on the length $L=40$ ring), in keeping with general expectations from the study of the delocalized impurity in the Tonks-Girardeau limit.\cite{GamayunPRE14,GamayunPRA14,LychkovskiyPRA2014} We have also examined the density of the impurity in the DE to ascertain whether translational symmetry is restored in the long-time limit. Generically,  we find that translational symmetry is not restored in the finite-size system due to a symmetry of the Bethe states under a change in sign of all the momenta and rapidities.

\textit{Conclusion}.---
In this letter, we consider the nonequilibrium time evolution of a single localized impurity~\fr{Eq:Initial_State} injected into the ground state of the Lieb-Liniger model. In both the case of zero and finite COM momentum, we observe a `stuttering' behavior in the motion. In the first case (see Fig.~\ref{Fig:Gaussian_Q0}), this quantum stutter manifests in the arrested expansion of the impurity (in the absence of a lattice). This arises from the out-of-equilibrium formation of quasibound impurity-hole pairs which are stable for extended periods of time.  This interaction-driven effect can be qualitatively captured by the MF decoupling~\fr{Eq:Mean_Field}: the impurity repels the background gas, leading to the formation of a hole which acts as a confining potential for the impurity. Eventually the impurity broadens in a sequence of rapid expansions and quasistationary periods, all the while undergoing small amplitude breathing oscillations. This stuttering motion and the quasibound state formation highlights the importance of distinguishability, as this mechanism does not exist for an impurity of the same species as the background gas.\footnote{We show results in this case in~[\onlinecite{SuppMat}].} 

In contrast, when the impurity is injected with a finite COM momentum, the quantum stutter is clearly seen in the motion of the impurity, which snakes through the background gas, see Figs.~\ref{Fig:Gaussian_Q}\ \&\ \ref{Fig:Center_of_mass}. We can picture this as a quantum Newton's cradle\cite{WeissNature06} on the ring: the injected impurity scatters particles in the background gas until it loses most of its COM momentum. These scattered excitations then propagate around the ring and subsequently collide with the impurity, causing it to move once again. This process repeats, leading to the stuttering, snaking motion of the COM. Quantum flutter, the exchange of momentum back-and-forth between a delocalized impurity and the background gas, has been studied in the Tonks-Girardeau regime.\cite{MathyNatPhys12,KnapPRL14,BurovskiPRA14,GamayunPRA14,GamayunPRE14,LychkovskiyPRA2014} The momentum of the impurity in the long-time limit was computed by means of the DE and found to be small, but non-zero.
   
Our results are of direct relevance to experiments in cold atomic gases and the observed physics should not be reliant upon the integrability of the model (see, e.g., Refs.~\ocite{KnapPRL14} and \ocite{SuppMat}) and should survive finite temperature.\cite{BoudjemaaPRA14} The discussed results may also be useful in elucidating the properties of the TCLLM at finite temperature, where it is likely that impurity-like solitons arise.\cite{KarpiukPRL12,KarpiukPRA15} Finally, this work provides a nontrivial check and validation of cutting-edge theoretical results for the matrix elements of the TCLLM in the extreme imbalance limit.\cite{PozsgayJPhysA12,PakuliakArxiv15}

{\textit{Acknowledgments}.---} 
We thank Fabian Essler, Bruno Bertini, John Goold, Rianne van den Berg and Giuseppe Brandino for useful discussions surrounding this work. This work was partially supported by the EPSRC under Grant No. EP/I032487/1 (NJR), IRSES Grant QICFT (NJR, RMK), the FOM and NWO foundations of the Netherlands (JSC, RMK), and the U.S. Department of Energy, Office of Basic Energy Sciences, under Contract Nos. DE-AC02-98CH10886 and DE-SC0012704 (NJR, RMK).
 
\bibliography{IntegrabilityBib}
 
\onecolumngrid 

\vspace{0.5cm}

\begin{center}
\textbf{\large Supplemental Materials for ``Motion of a distinguishable impurity in the Bose gas: Arrested expansion without a lattice and impurity snaking''}
\end{center}

\vspace{0.5cm}

\twocolumngrid

\setcounter{equation}{0}
\setcounter{figure}{0}
\setcounter{table}{0}
\makeatletter
\renewcommand{\theequation}{S\arabic{equation}}
\renewcommand{\thefigure}{S\arabic{figure}}

\subsection{Summary of required matrix elements for the two-component Lieb-Liniger model}
\label{App:MatrixElements}

Here we summarize known results\cite{PozsgayJPhysA12} for matrix elements of local operators in the two-component Lieb-Liniger model. For technical reasons, the known matrix elements of Ref.~\ocite{PozsgayJPhysA12} are restricted to the case with a single impurity boson ($N_1 = 1$) and to the local operators $\Psi_1(0)$, $\Psi\dg_1(0)\Psi_1(0)$ and $\Psi\dg_1(0)\Psi_2(0)$. As a consequence of this, we are not able to `image' the background gas $\Psi\dg_2(0)\Psi_2(0)$. Very recently\cite{PakuliakArxiv15} there have been new (and more general) results for matrix elements in the two-component Lieb-Liniger model, but we have yet implement them. 

\subsubsection{Normalization conventions} 

We focus on $N$-particle eigenstates containing a single $N_1$ boson. For clarity and ease of comparison, here we work with the conventions of Ref.~\ocite{PozsgayJStatMech11} and we define the (non-normalized) eigenstates $|\{q\}_N;\mu\ra\ra$. The eigenstates in the body of the text are recovered by normalization $|\{q\}_N;\mu\ra = |\{q\}_N;\mu\ra\ra/ ||  \{q\}_N;\mu ||$, where norms of the eigenstates are given by 
\bea
&&||\{q\}_N;\mu||^2 = \la\la \{q\}_N;\mu | \{q\}_N;\mu\ra\ra = c\ {\rm det} {\cal J}_2,\nn
&&{\cal J}_2 = \left( \begin{array}{cc} J_{qq} & J_{q\mu} \\ J_{\mu q} & J_{\mu\mu} \end{array}\right).\nonumber
\eea
Here ${\cal J}_2$ is the Jacobian of the nested Bethe ansatz equations [see Eqs.~(2,3)] given by the matrix elements
\bea
(J_{qq})_{jl} &=& \delta_{jl} \Bigg[ L + \sum_{m=1}^N \varphi_1(q_j-q_l) - \varphi_2(q_m - \mu)\Bigg]\nn
&&- \varphi_1(q_j - q_l),\nn
(J_{q\mu})_{j1}&=& (J_{\mu q})_{1j} = \varphi_2(k_j-\mu),\nn
J_{\mu\mu} &=& \sum_{m=1}^N \varphi_2(k_m - \mu),\nonumber
\eea
where we define the scattering phase $\varphi_n(u) = 2cn/(n^2u^2+c^2)$ and $j,l = 1,\ldots,N$. 

We take the one-component Lieb-Liniger eigenstates $|\{p\}_N\ra\ra$ to have their conventional normalization\cite{KorepinBook,ReshetikhinJSovMath89,PangJMathPhys90,GohmannPhysLettA99}
\bea
||\{p\}_N||^2 &=& \prod_{j<l}\Big[ (p_j - p_l)^2 + c^2\Big]{\rm det}{\cal J}_1,  \nn
({\cal J}_1)_{jl} &=& \delta_{j,l} \Bigg[L + \sum_{m=1}^N \varphi_1(p_j-p_m)\Bigg] - \varphi_1(p_j-p_l).\nonumber
\eea

\subsubsection{Matrix elements $\la\la \{p\}_N | \Psi_1(0) | \{k\}_N;\lambda\ra\ra$}
For two states with no coinciding momenta, the matrix element of the impurity annihilation operator takes the determinant form
\bea
&&\la\la \{p\}_{N-1} | \Psi_1(0) |\{k\}_N;\lambda\ra\ra\nn
&&= \frac{\prod_{i>j}( k_i - k_j + ic )}{\prod_{l>m} (p_l - p_m + ic)}\frac{-ic}{\prod_j (\lambda-k_j -ic/2)}{\rm det}\mathbb{M}.\nonumber
\eea
Here the $(N-1)\times(N-1)$ matrix $\mathbb{M}$ has elements $\mathbb{M}_{jk} = M_{jk} - M_{N,k}$ with
\bea
M_{jk} &=& t(p_k - k_j)h_2(\lambda-k_j) \frac{\prod_{m=1}^{N-1} h_1(p_m-k_j)}{\prod_{m=1}^{N} h_1(k_m-k_j)}\nn
&& +  t(k_j-p_k)h_2(k_j-\lambda) \frac{\prod_{m=1}^{N-1} h_1(k_j-p_m)}{\prod_{m=1}^{N} h_1(k_j-k_m)}\nonumber
\eea
where we've defined the functions $h_n(u) = u + ic/n$ and $t(u) = -c/[u(u+ic)]$. 

\subsubsection{Matrix elements $ \la\la\{p\}_N;\mu|\Psi_1\dg(0)\Psi_1(0)|\{k\}_N;\lambda\ra\ra$}
For two states with no coinciding momenta, the matrix elements of the impurity density operator are
\bea
&&\la\la\{p\}_N;\mu|\Psi_1\dg(0)\Psi^{\phantom\dagger}_1(0)|\{k\}_N;\lambda\ra\ra \nn
&&= \frac{-i}{c}(-1)^{N(N+1)/2}\prod_{j>l}\frac{1}{k_j - k_l -ic} \prod_{j>l}\frac{1}{p_j-p_l+ic}{\rm det} \mathbb{V} \nn
&&~\times \prod_{l,m=1}^{N} \Big(k_l-p_m+ic\Big) \frac{c^2}{\prod_j (\lambda-k_j - ic/2)(\mu-p_j+ic/2)}\nonumber
\eea
where the $(N+1)\times(N+1)$ matrix $\mathbb{V}$ has elements
\bea
\mathbb{V}_{jl} &=& \bigg(p_l-\lambda+\frac{ic}{2}\bigg)\bigg(p_l-\mu-\frac{ic}{2}\bigg)\tilde t(k_j-p_l) \nn
&& + \bigg(p_l-\lambda-\frac{ic}{2}\bigg)\bigg(p_l-\mu+\frac{ic}{2}\bigg)\tilde t(p_l-k_j)\nn
&& \times \prod_{m=1}^{N}\frac{(p_l-k_m+ic)(p_l-p_m-ic)}{(p_l-k_m-ic)(p_l-p_m+ic)} \nn
\mathbb{V}_{N+1,j} &=& \prod_{m=1}^N \frac{p_m-p_j + ic}{k_m-p_j + ic},\qquad \mathbb{V}_{j,N+1}=1,\nn
\mathbb{V}_{N+1,N+1}&=&0, \nonumber
\eea
with $j,l=1,\ldots,N$.
Additionally, the diagonal elements of the impurity density operator follow from translational invariance of the eigenstates: $\la \{k\}_N;\lambda|\Psi\dg_1(0)\Psi_1(0)|\{k\}_N;\lambda\ra = 1/L$. 
 
\subsection{Dynamics of an indistinguishable impurity in the one-component Lieb-Liniger model}

Here we consider the initial state 
\be
|\Psi_2(Q)\ra = \frac{1}{{\cal N}_2} \int\limits_0^L\rd x\ e^{iQx} e^{-\frac12\left(\frac{x-x_0}{a_0}\right)^2} \Psi\dg_2(x)|\Omega\ra,
\label{Eq:Initial_State_onecomp}
\ee
where $|\Omega\ra$ is the ground state of $N_2$ bosons of species $2$, e.g. the analogue of Eq.~(4) with an indistinguishable impurity. We wish to consider the time evolution of the expectation value of the density operator $\Psi\dg_2(x)\Psi_2(x)$ with this initial state. Our prescription for computing the time evolution is analogous to the two-component case: we expand the expectation value $\rho_2(x,t) = \la \Psi_2(Q,t)|\Psi\dg_2(x)\Psi_2(x)|\Psi_2(Q,t)\ra$ in terms of known matrix elements and overlaps between the Bethe states,\cite{ReshetikhinJSovMath89,PangJMathPhys90,SlavnovThMathPhys90,KojimaCommMathPhys97,KorepinIntJModPhysB99,GohmannPhysLettA99,KorepinBook,PozsgayJStatMech11} to obtain an expansion similar to Eq.~(5). 

For the case with $Q=0$, we can gain some insight from examining the noninteracting limit $c=0$. Working on the infinite system ($L\to \infty$) with $x_0 = 0$, we find the density 
\bea
\rho_2(x,t) &=&  \rho +  \frac{a_0^2}{\sqrt{\pi a_0^2} + 2\pi a_0^2 \rho}\left[ \frac{e^{-\frac{a_0^2 x^2}{a_0^4 + 4t^2}}}{\sqrt{a_0^4 + 4t^2}}\right. 
\nn
&&+ \left. \rho \sqrt{2\pi} \frac{e^{-\frac{a_0^2}{2}\frac{x^2}{a_0^4+4t^2}}}{(a_0^4+4t^2)^{\frac14}}
\cos\left(\theta_t - \frac{tx^2}{a_0^4 + 4t^2}\right)\right],\nn \label{Eq:Nonint}
\eea
where $\rho$ is the average density and with $\rho$ and $2\theta_t = \arctan(2t/a_0^2)$. Thus we expect the wave packet to be of Gaussian shape with oscillations superimposed on top. 

\begin{figure}
\includegraphics[width=0.45\textwidth]{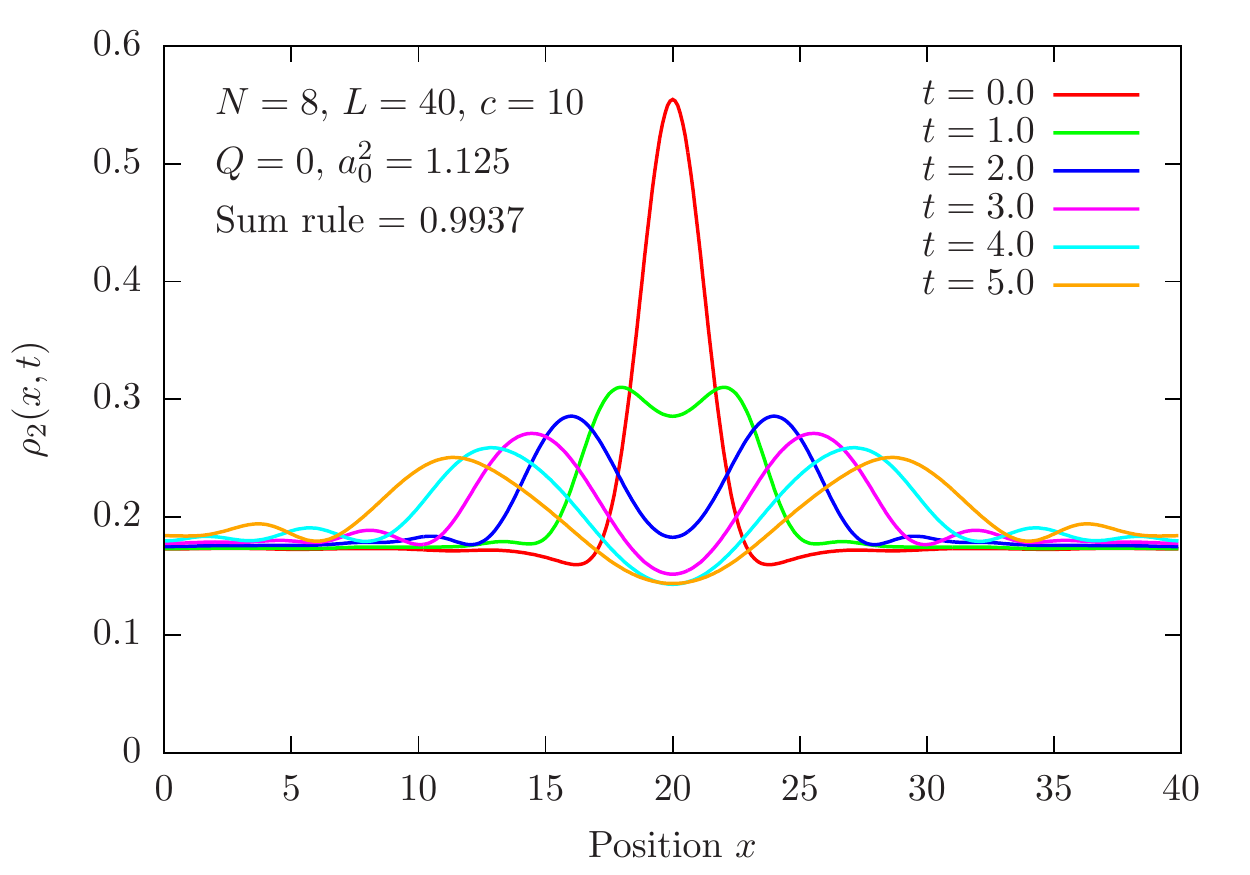} 
\caption{Time evolution of the density profile $\rho_2(x,t)$ for the initial state~\fr{Eq:Initial_State_onecomp} in the one-component 
Lieb-Liniger model for $N=8$ particles on the circumference $L=40$ ring with interaction parameter $c=10$. We used 
12619 states to saturate the sum rule to 0.9937.}
\label{Fig:Gaussian_OneComp}
\end{figure}

We present the time evolution of the density operator on the initial state~\fr{Eq:Initial_State_onecomp} for $N=8$ particles, $Q=0$, $a_0^2 = 1.125$ and $L=2x_0=40$ in Fig.~\ref{Fig:Gaussian_OneComp}. We 
see that the behavior of the density as it evolves in time is qualitatively consistent with the noninteracting result~\fr{Eq:Nonint}. We see no evidence of a stalling of the spreading of the wave packet, which is not
surprising as the mechanism which exists in the two-component case (hole formation in the background gas and subsequent trapping of the impurity in the hole) is not present when there is a single species of boson.  

\subsection{Lattice mean field description}
\label{App:MF}

In an attempt to explain the dynamics of the initial state, Eq.~(4) in the main text, we consider the following lattice Hamiltonian (we consider a lattice Hamiltonian for
numerical convenience):
\bea
H_{\rm latt} &=& -J_d \sum_l \Big( d\dg_l d^{\phantom\dagger}_{l+1} + {\rm H.c}\Big)+ U \sum_l \Big( n^b_l + n^d_l\Big)^2 \nn
&& - J_b \sum_l \Big(b\dg_l b^{\phantom\dagger}_{l+1} + {\rm H.c.}\Big),
\label{Eq:MF_Ham}
\eea
where $n^a_l = a\dg_l a_l$ is the number operator. This model is motivated by the two-component Lieb-Liniger model: we consider two species of bosons which have an on-site interaction only and the kinetic terms coincide in the continuum limit
\bea
\lim_{a_0\to 0 } d(x)\dg d(x+a_0) &\to& d\dg(x) d(x) + a_0 d\dg(x) \p_x d(x)\nn && + \frac{a_0^2}{2} d\dg(x) \p_x^2 d(x), \nn
\sum_l d\dg_l d^{\phantom\dagger}_{l+1} + d\dg_l d^{\phantom\dagger}_{l-1} &\to& {\rm const. } + a_0 \int {\rd} x \ d\dg(x) \p_x^2 d(x). \nonumber
\eea
We choose the $d$ bosons to play the role of the background gas (species $2$ in the main body of the manuscript) and we consider $U\gg J_d$ to reflect the strong coupling regime of the main text.

The Heisenberg equations of motion for the boson bilinears take the form
\begin{widetext}
\bea
\frac{\rd}{\rd t} b\dg_ib^{\phantom\dagger}_j &=& 
-iJ_b \Big( b\dg_{i-1}b^{\phantom\dagger}_j + b\dg_{i+1}b^{\phantom\dagger}_j - b\dg_ib^{\phantom\dagger}_{j+1} - b\dg_ib^{\phantom\dagger}_{j-1}\Big)
+ iU\Big[ b\dg_i b^{\phantom\dagger}_j ( n_i^b - n_j^b) + ( n_i^b - n_j^b)b\dg_i b_j+ 2(n^d_i - n^d_j) b\dg_i b^{\phantom\dagger}_j \Big] \ ,\nonumber
\eea
and similarly for $d\dg_i d_j$ with $d\leftrightarrow b$. We take the expectation value of this expression and perform a time-dependent mean-field decoupling which preserves the $U(1)$ symmetry for each of 
the species (cf. Eq.~(7))
\bea
&&\la (n^d_i-n^d_j) b\dg_i b^{\phantom\dagger}_j \ra \to \Big(\la n^d_i(t) \ra - \la n^d_j(t)\ra\Big) \la b\dg_i(t) b^{\phantom\dagger}_j(t)\ra, \qquad 
\la n^b_i (t) b\dg_i(t) b^{\phantom\dagger}_j(t) \ra \to \la n^b_i(t) \ra \la b\dg_i(t) b^{\phantom\dagger}_j(t)\ra + \la b\dg_i b^{\phantom\dagger}_j(t)\ra\la b^{\phantom\dagger}_i b\dg_i(t)\ra, \nonumber
\eea
to arrive at the \textit{approximate} equations of motion for the boson bilinears:
\bea
\frac{\rd}{\rd t} \la b\dg_ib^{\phantom\dagger}_j(t)\ra &=& 
-iJ_b\Big[ \la b\dg_{i-1}b^{\phantom\dagger}_j(t)\ra - \la b\dg_i b^{\phantom\dagger}_{j+1}(t)\ra + \la b\dg_{i+1}b^{\phantom\dagger}_j(t)\ra - \la b\dg_ib^{\phantom\dagger}_{j-1}(t)\ra \Big]\nn
&& +2iU \Big[ \la n^d_i(t)\ra - \la n^d_j(t)\ra + 2 \la n^b_i(t) \ra - 2 \la n^b_j(t)\ra\Big] \la b\dg_ib^{\phantom\dagger}_j(t)\ra, \label{Eq:MF_EoM}
\eea
\end{widetext}
with similar for the $d$ bosons. Our initial conditions are fixed by the initial state $|\Psi(Q)\ra$; for the purposes of convenience, we consider the ground state $|\Omega\ra$ to be the $c=0$ ground state (this is an approximation, alternatively one can view this situation as a combination of injecting the impurity and performing a quantum quench of the interaction parameter), in which only the zero-mode is populated. The initial conditions for the bilinears are: 
\be
\begin{aligned}
\la b\dg_i b^{\phantom\dagger}_j \ra_0 &= \frac{1}{|{\cal N}|^2} e^{-\frac12 \left(\frac{i-i_0}{a_0}\right)^2} e^{-\frac12 \left(\frac{j-i_0}{a_0}\right)^2} e^{iQ(i-j)},\\
 \la d\dg_i d^{\phantom\dagger}_j\ra_0 &= \rho,
\end{aligned}
\label{Eq:MF_Initial_Conditions}
\ee
where $\rho = N/L$ is the density of the background gas and $|{\cal N}|^2 = \sum_x \exp[ -( x-x_0 )^2/a_0^2]$ is a normalization factor.

\begin{figure}[ht]
\begin{tabular}{l}
(a) \\
\includegraphics[trim=140 70 40 125,clip,width=0.38\textwidth]{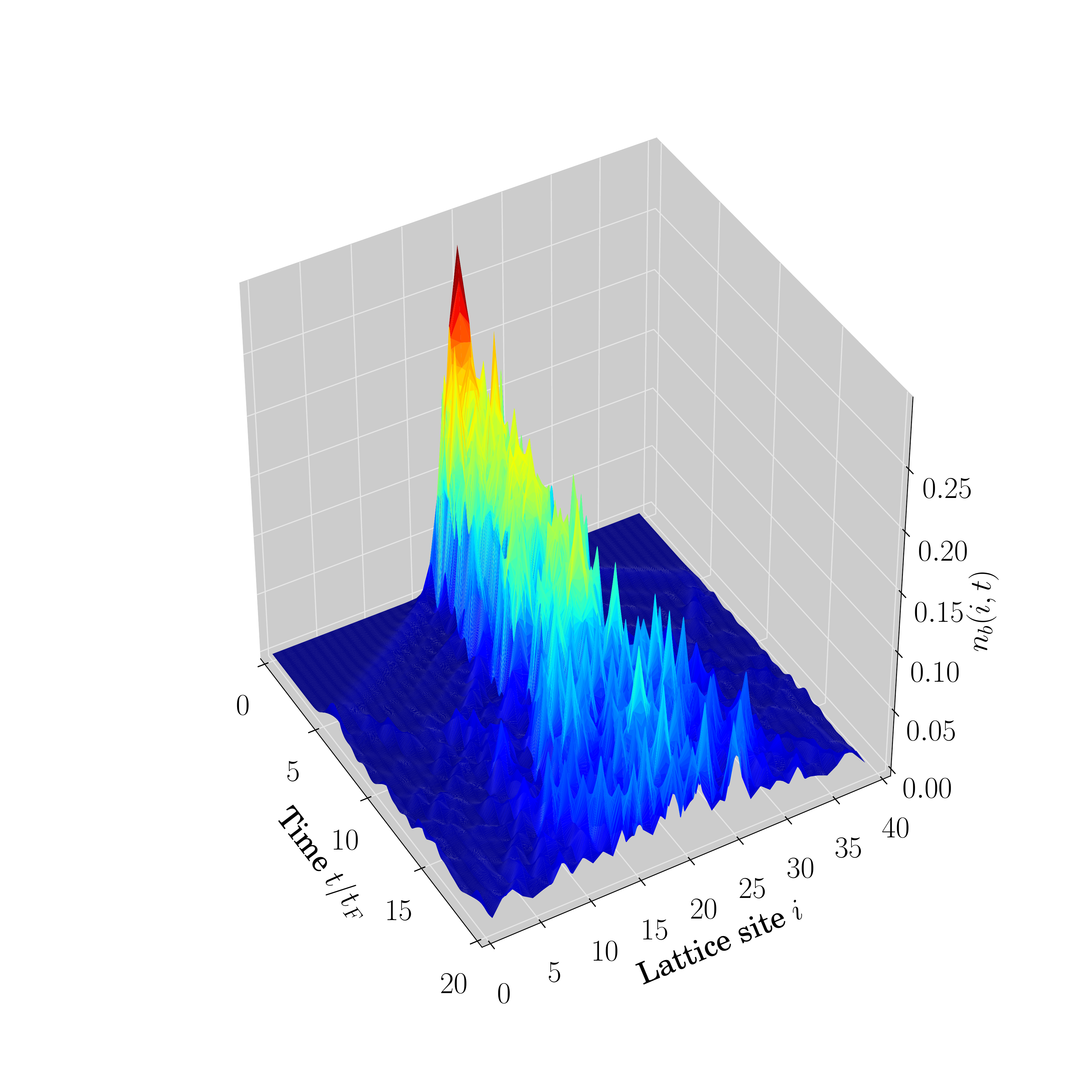} \\
(b) \\ 
\includegraphics[trim=140 70 40 125,clip,width=0.38\textwidth]{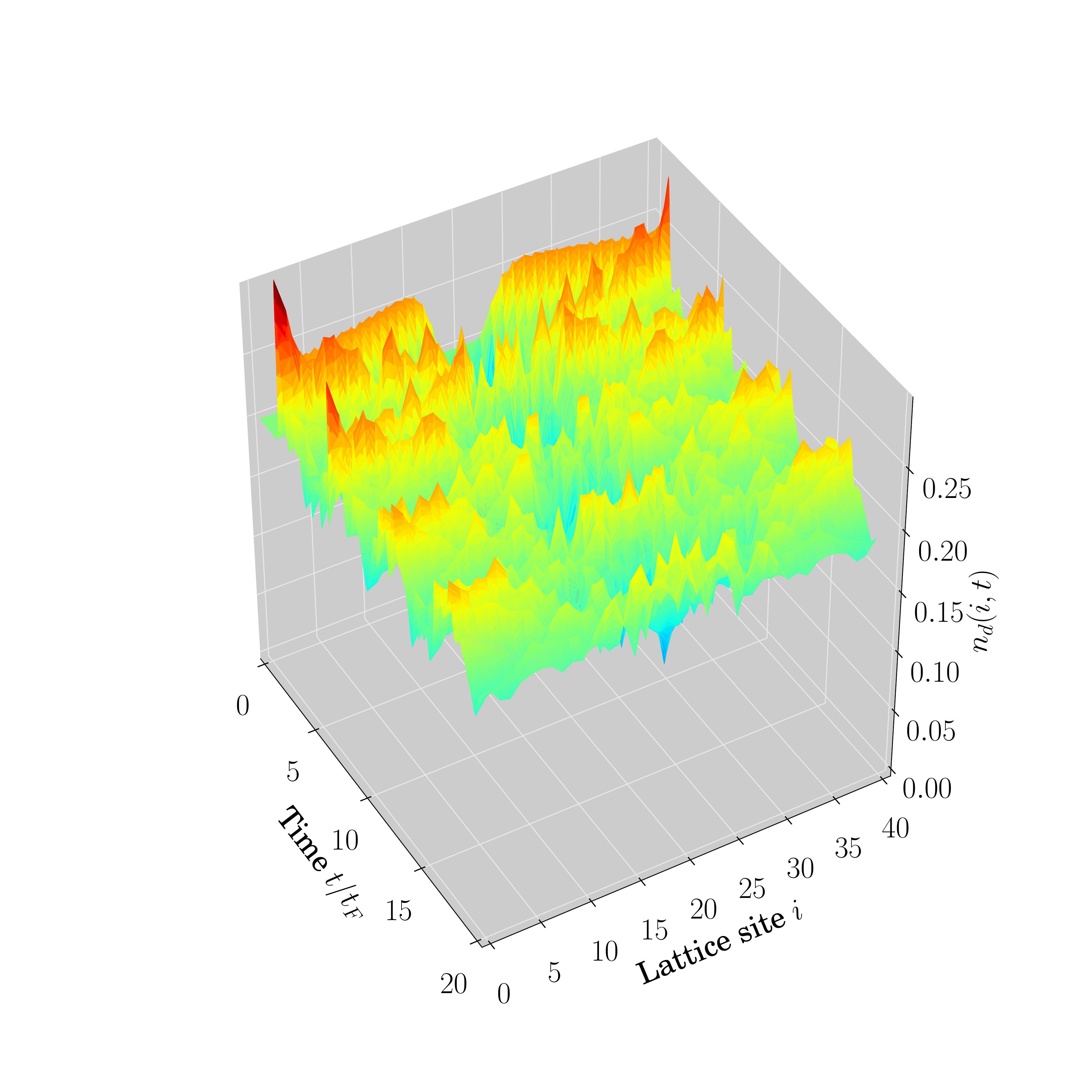} 
\end{tabular}
\caption{Time depedence of the (a) $n_b$; (b) $n_d$ boson number operator expectation values from the mean-field equations of motion~\fr{Eq:MF_EoM} with $U=14.5$ and $J_d = J_b = 1$. Initial conditions~\fr{Eq:MF_Initial_Conditions} with $a_0=2$, $Q=0$ and $\rho = 0.2$ for $L=40$ sites were used. }
\label{Fig:MFa}
\end{figure}

\begin{figure}[ht]
\begin{tabular}{l}
(a) \\
\includegraphics[trim=140 70 40 125,clip,width=0.38\textwidth]{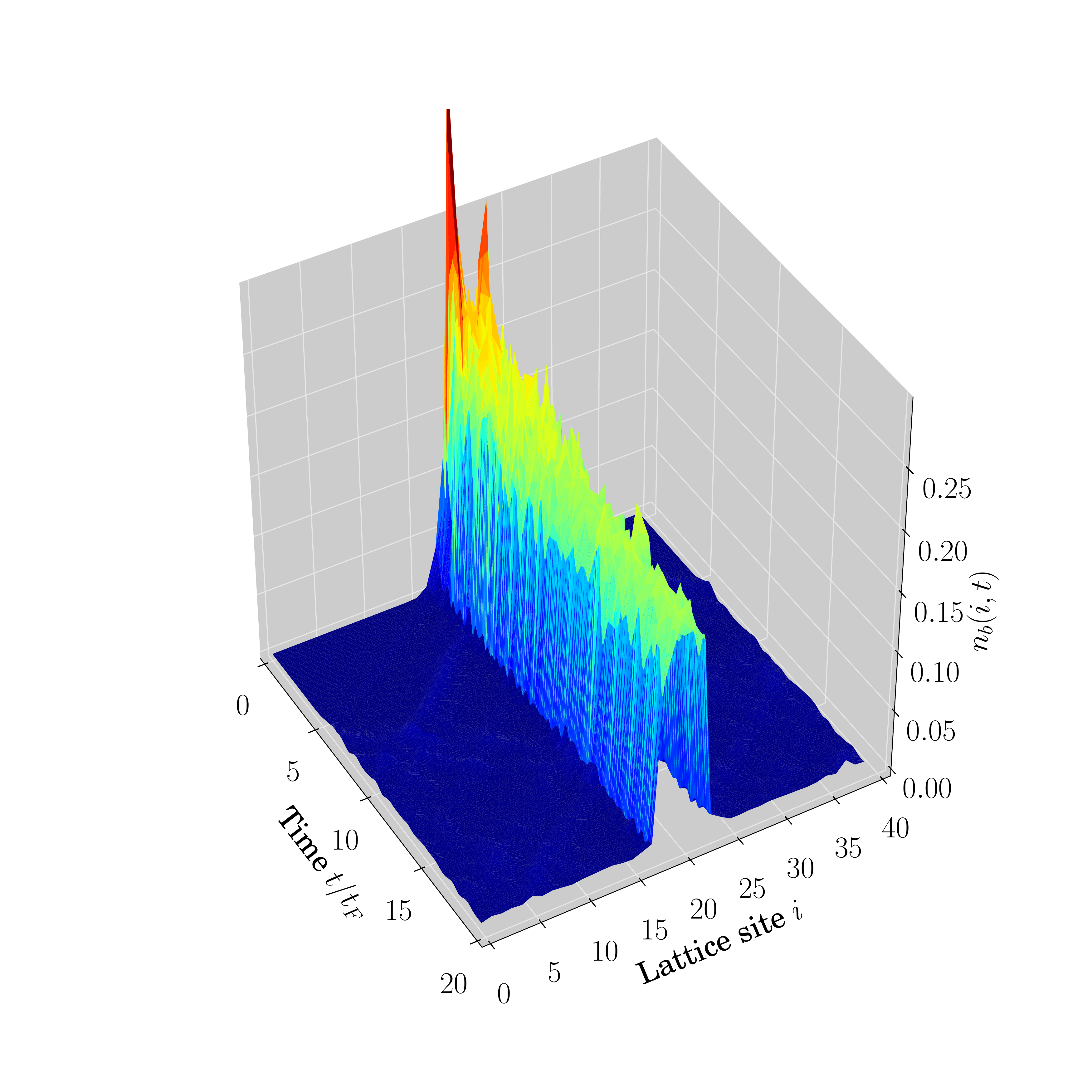}  \\
(b) \\ 
\includegraphics[trim=140 70 40 125,clip,width=0.38\textwidth]{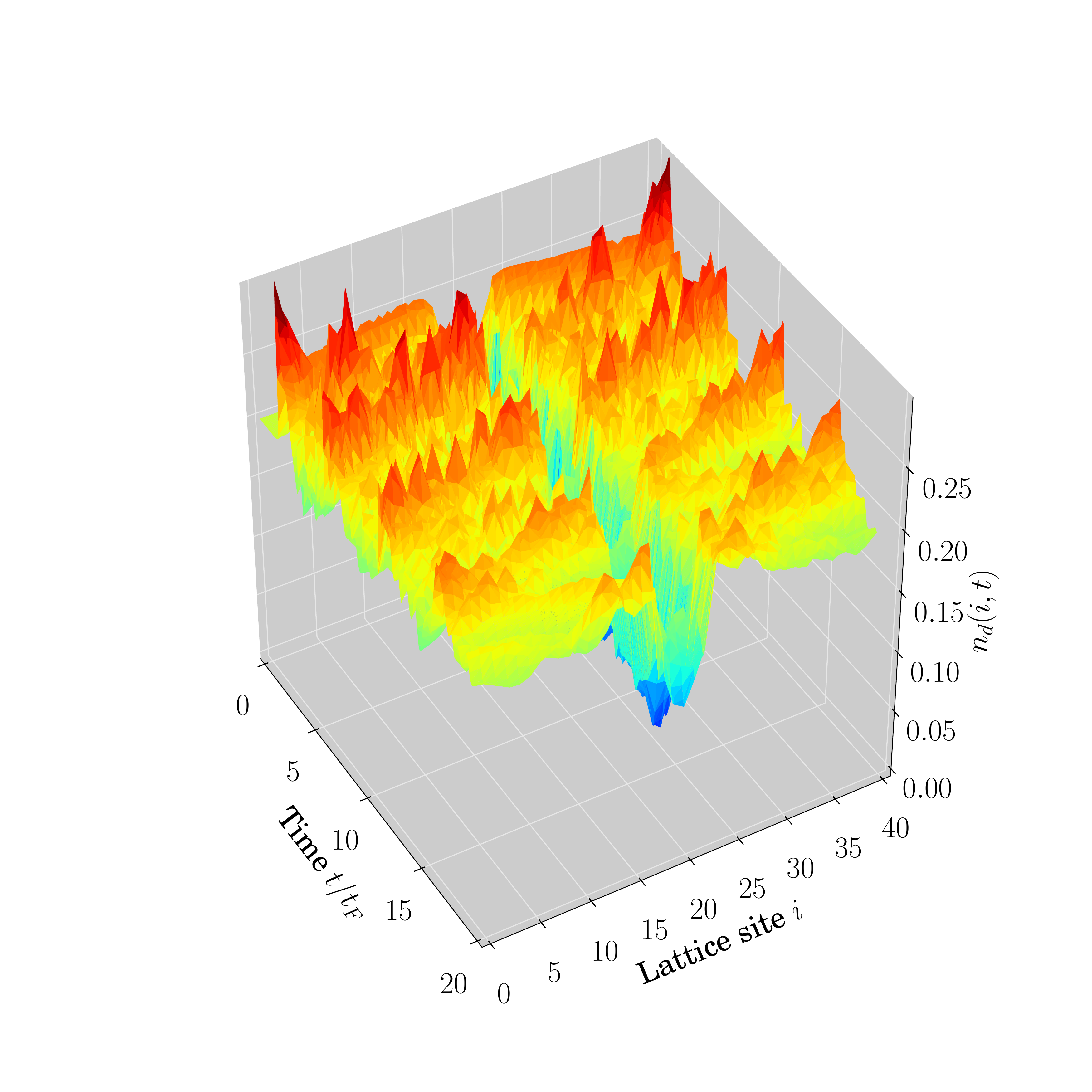} 
\end{tabular}
\caption{Time depedence of the (a) $n_b$; (b) $n_d$ boson number operator expectation values from the mean-field equations of motion~\fr{Eq:MF_EoM} with $U=14.5$ and $J_d = J_b = 1$. Initial conditions~\fr{Eq:MF_Initial_Conditions} with $a_0=2$, $Q=13 \times 2\pi/L$ and $\rho = 0.2$  for $L=40$ sites were used. .}
\label{Fig:MFb}
\end{figure}

We present results for the expectation values of the density operators in Figs.~\ref{Fig:MFa}~and~\ref{Fig:MFb} for $U=14.5$ and $J_d=J_b=1$ with $\rho = 0.2$ and $a_0=2$ on the circumference $L=40$ ring. Parameters were chosen in an attempt to qualitatively reproduce the $Q=0$ continuum behavior: arrested expansion followed by eventual spreading of the impurity ($cf.$ Fig.~\ref{Fig:Q0}). In Fig.~\ref{Fig:MFa} we see approximately the required behavior for $Q=0$: the impurity initially spreads, but for times $t\sim t_F - 9t_F$ expansion is arrested (the nature of the amplitude and fluctuations is clearly very different in the mean field lattice case compared to the continuum) before subsequently spreading. In the background gas, Fig.~\ref{Fig:MFa}(b), we see that a region of depleted density (a `hole') appears below the impurity, which remains despite multiple collisions with propagating wave packets (the red peaks crisscrossing the figure). 

In Fig.~\ref{Fig:MFb} we present similar data for the case with $Q=13\times(2\pi/L)$ (we move away from $Q=\pi$ as this is a special point in the lattice case). Surprisingly we see that the addition of finite center of mass momentum for the impurity has lead to a \textit{strengthening} of the dynamical arrest in the mean field approximation, with a deep and more robust hole forming in the background gas. Clearly we see no evidence of the `snaking' behavior observed in the continuum, despite multiple collisions with excitations in the background gas. This strongly suggests that the behavior observed in the continuum for $Q\neq0$ is beyond mean field theory (and may differ dramatically to that observed on the lattice).

\subsection{Addition plots: $Q=0$ long time and $Q=0,\pi$ constant-time cuts}

\begin{figure*}[ht]
\begin{tabular}{ m{0.5\textwidth} m{0.5\textwidth}}
(a) & (b) \\ 
\includegraphics[trim=200 110 75 180,clip,width=0.48\textwidth]{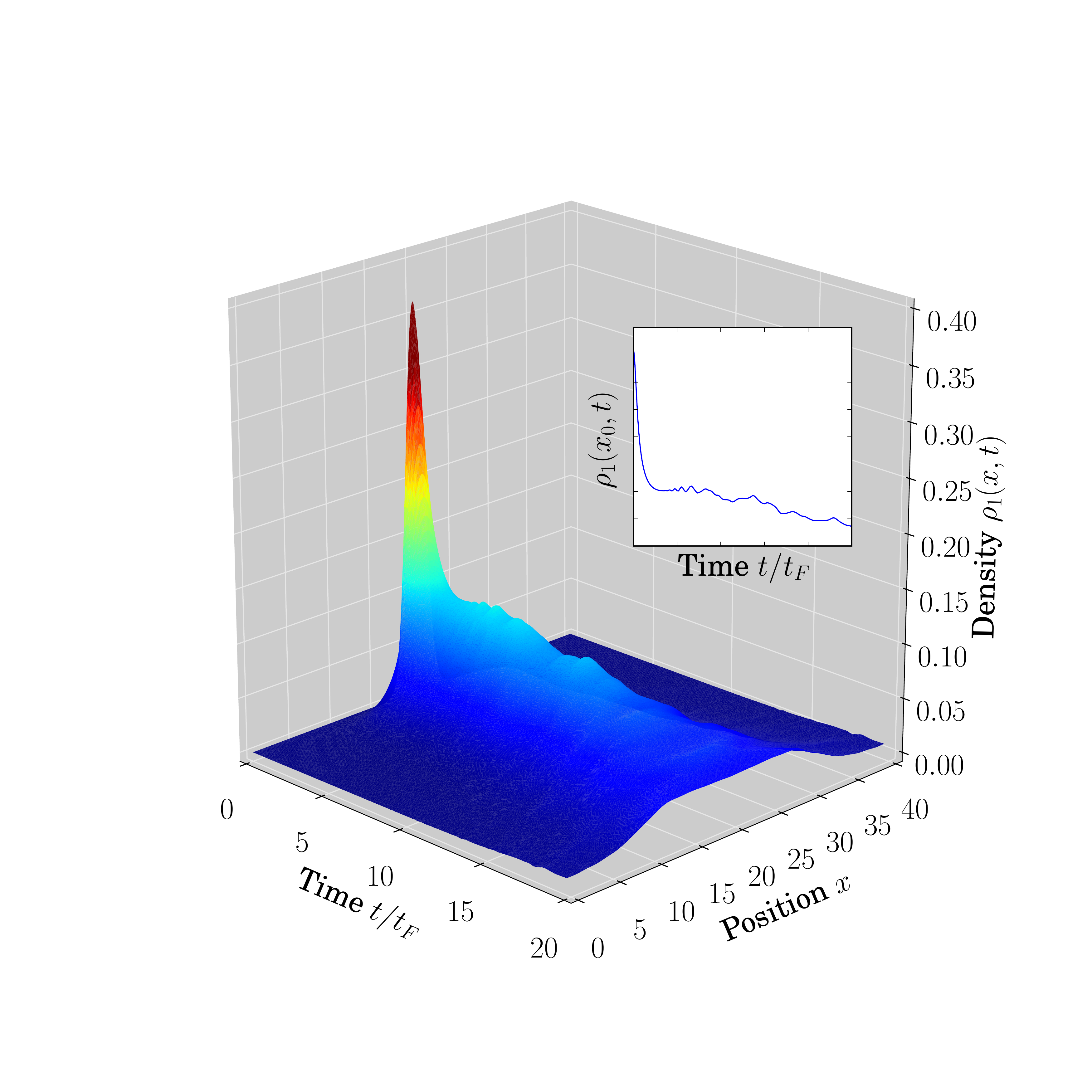} &
\begin{tabular}{l}\includegraphics[width=0.43\textwidth]{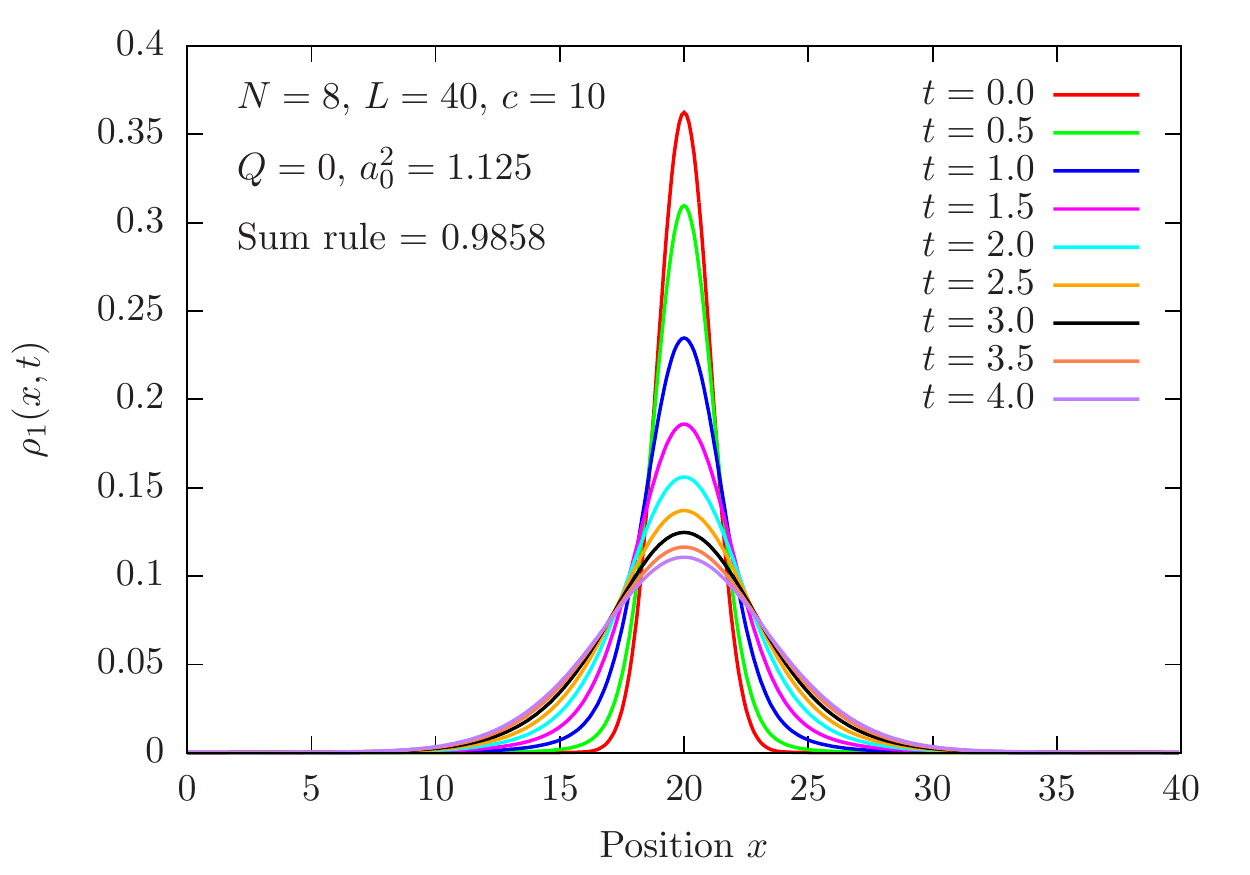} \\ \hspace{-1.75mm}\includegraphics[width=0.44\textwidth]{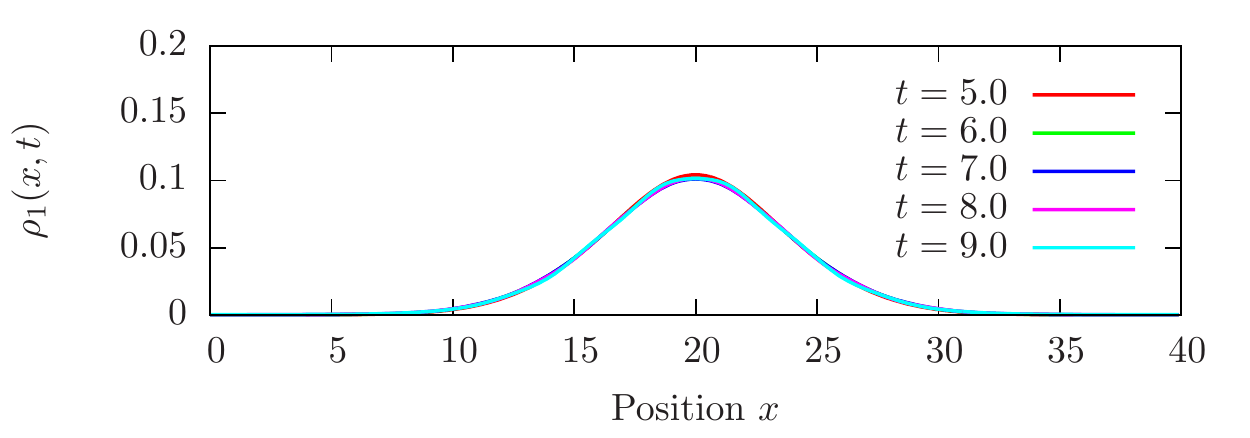} 
\end{tabular}
\end{tabular}
\caption{Time evolution of the impurity density of the initial state $|\Psi(Q)\ra$ [defined in Eq.(4)] with $Q=0$, $x_0=L/2$ and $a_0^2 = 1.125$ on the $L=40$ ring for a system of $N=8$ particles with interaction parameter $c=10$. The Hilbert space is truncated to 25150 states, leading to the sum rule [Eq.~(6)] $\ =0.9858$. (a) The full time-evolution, showing the initial rapid expansion, followed by a period of arrested expansion and subsequent spreading/quasi-stationary periods. Inset is the time-evolution of  the density at the midpoint. (b) Constant-time cuts at short times, showing in detail the rapid initial expansion and the period of arrested expansion. At intermediate times $5 \lesssim t \lesssim 18$ ($2t_F \lesssim t \lesssim 7t_F$) the impurity is approximately stationary and Gaussian in shape.}
\vspace{-0.25cm}
\label{Fig:Q0}
\end{figure*}

\begin{figure*}
\begin{tabular}{ccc}
$c = 5$ & $c=10$ & $c=20$ \\ 
\includegraphics[width=0.33\textwidth]{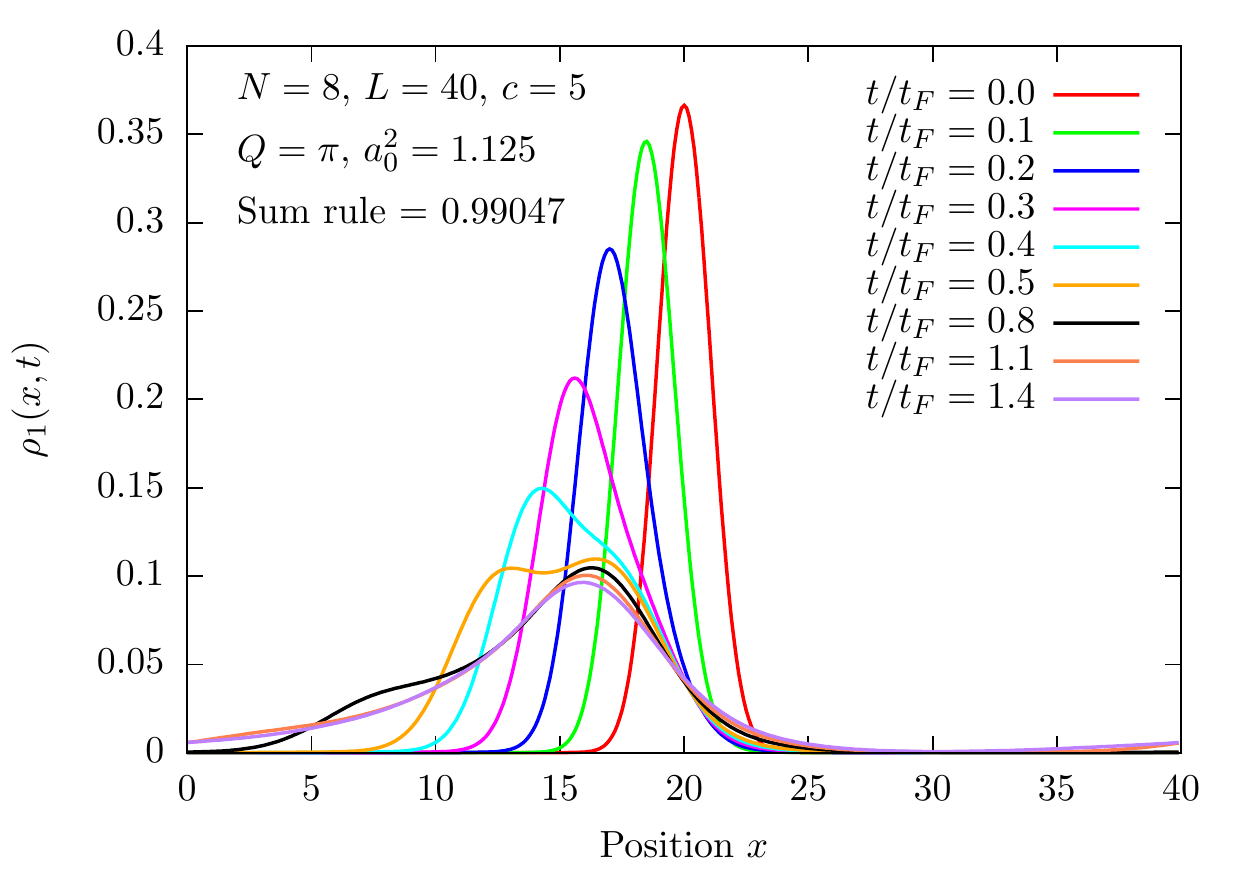} & 
\includegraphics[width=0.33\textwidth]{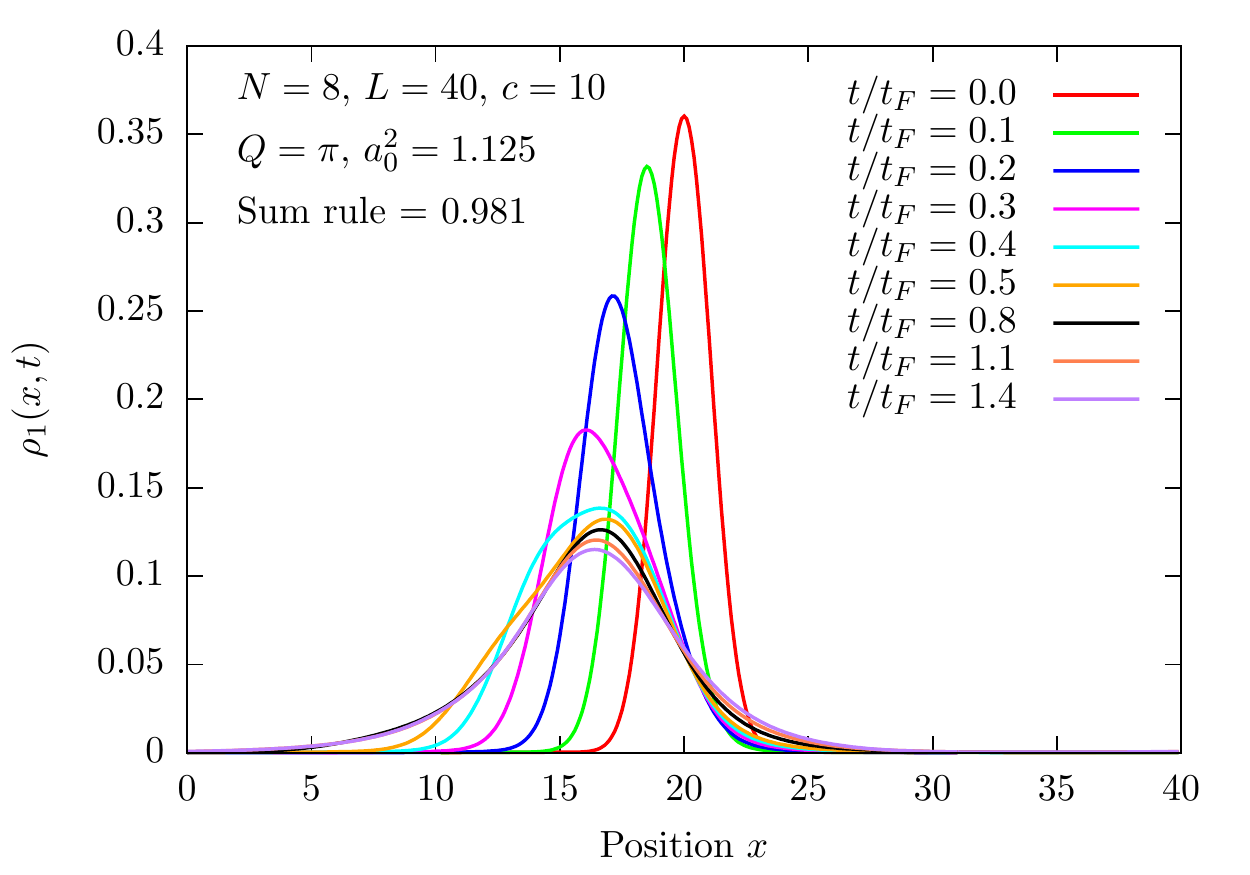} & 
\includegraphics[width=0.33\textwidth]{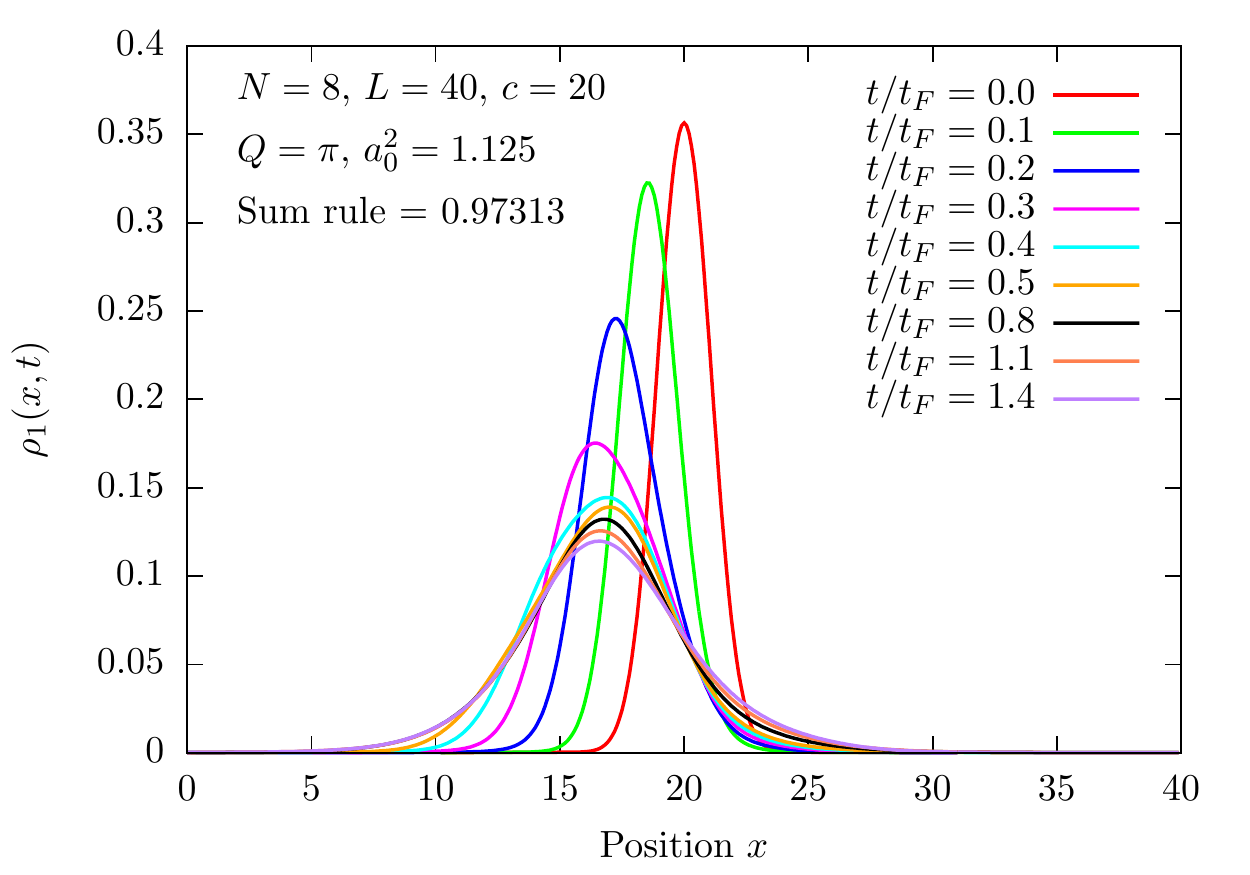} \\
\includegraphics[width=0.33\textwidth]{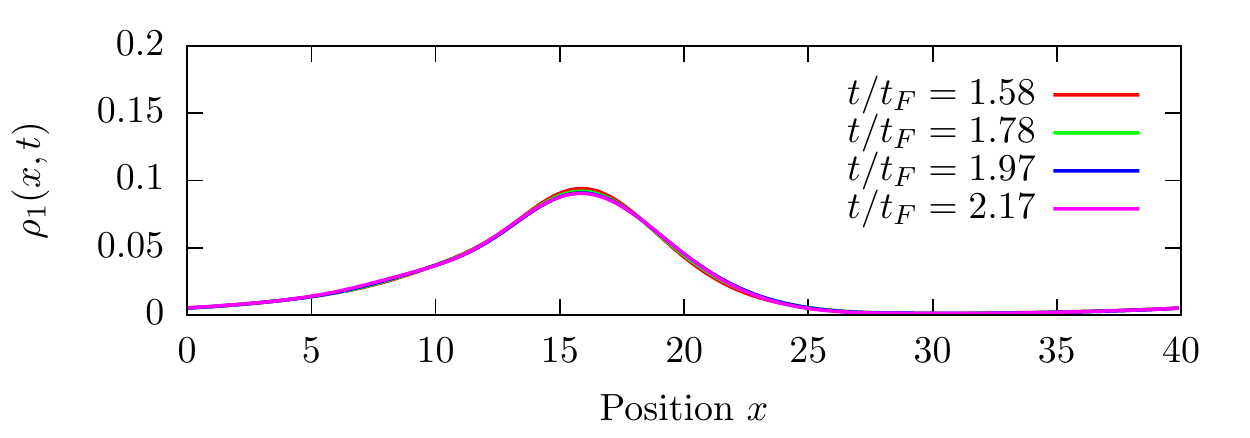} & 
\includegraphics[width=0.33\textwidth]{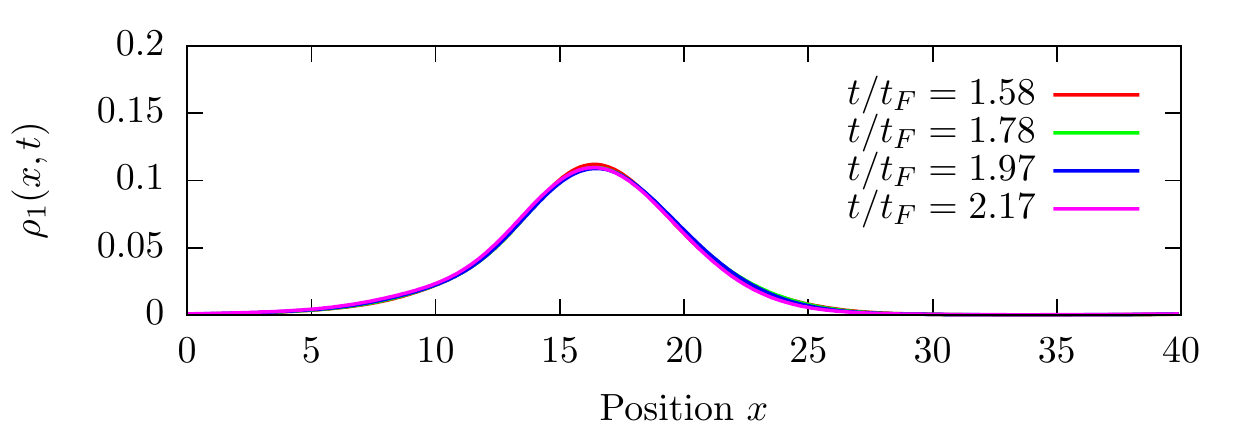} & 
\includegraphics[width=0.33\textwidth]{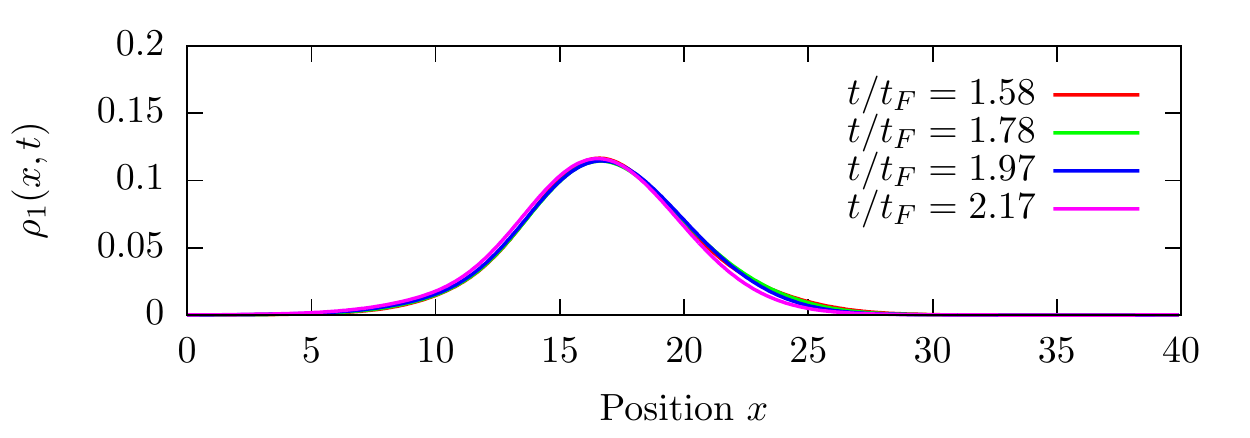} \\
\includegraphics[width=0.33\textwidth]{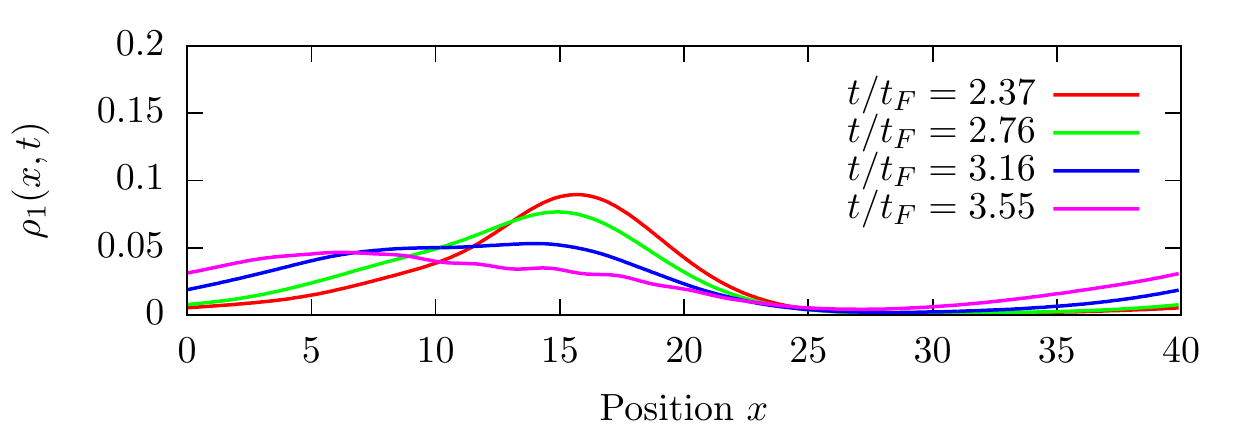} & 
\includegraphics[width=0.33\textwidth]{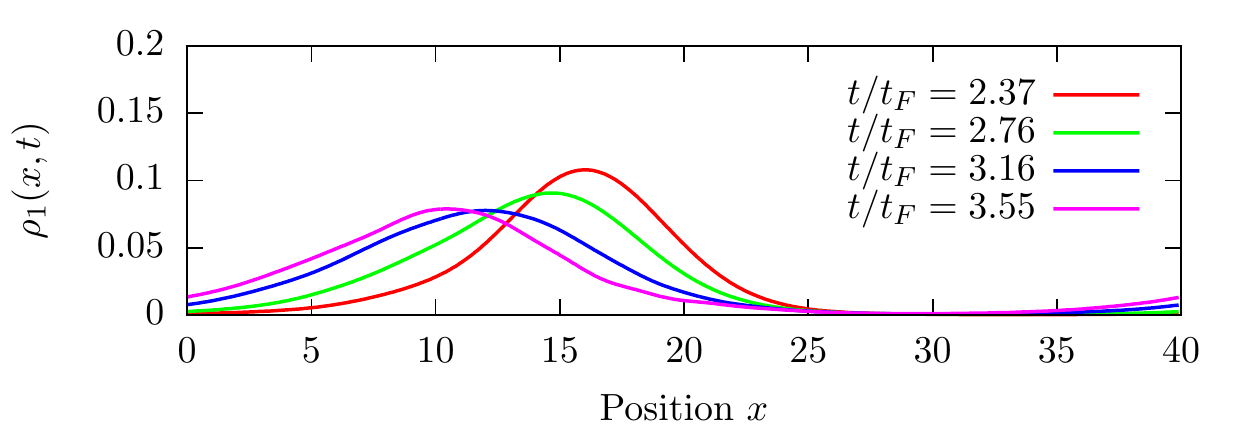} & 
\includegraphics[width=0.33\textwidth]{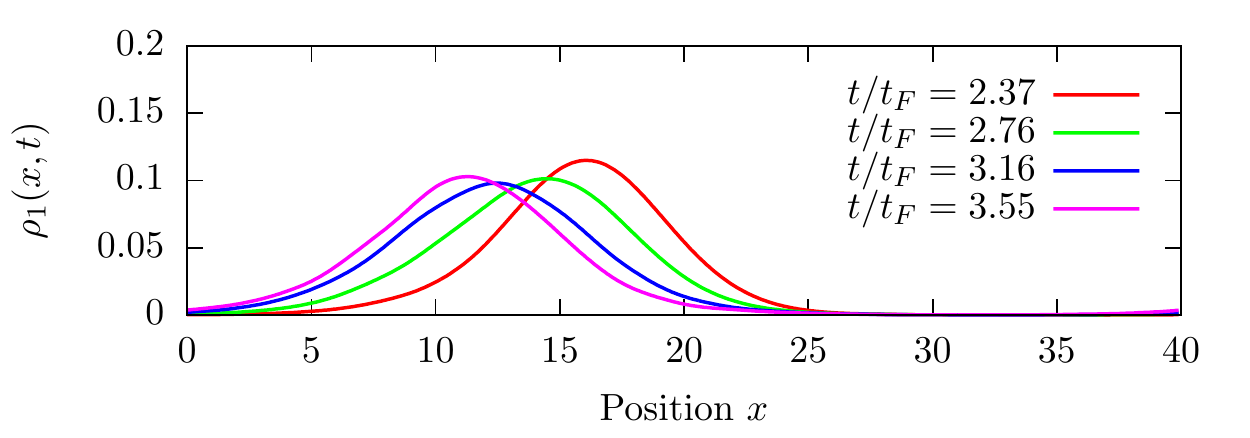} \\
\includegraphics[width=0.33\textwidth]{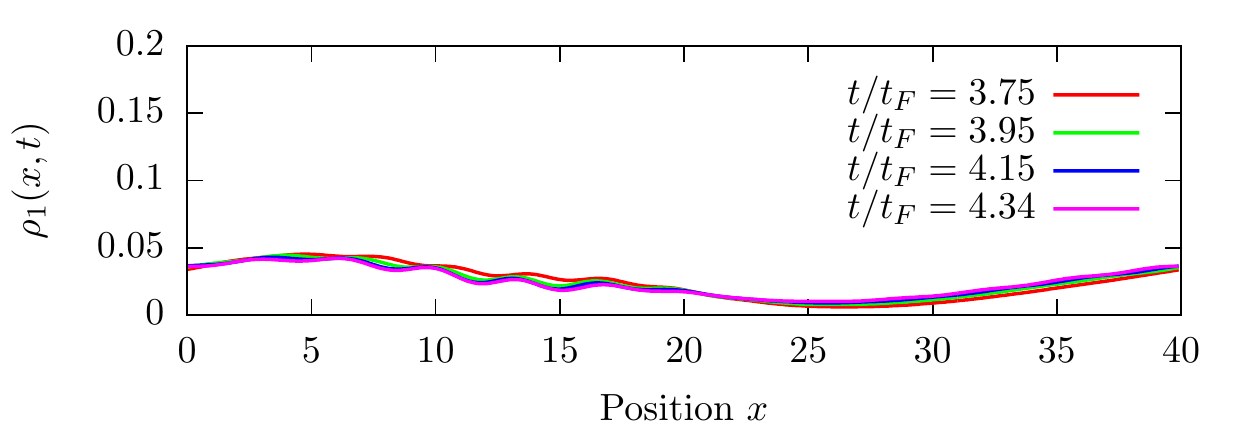} & 
\includegraphics[width=0.33\textwidth]{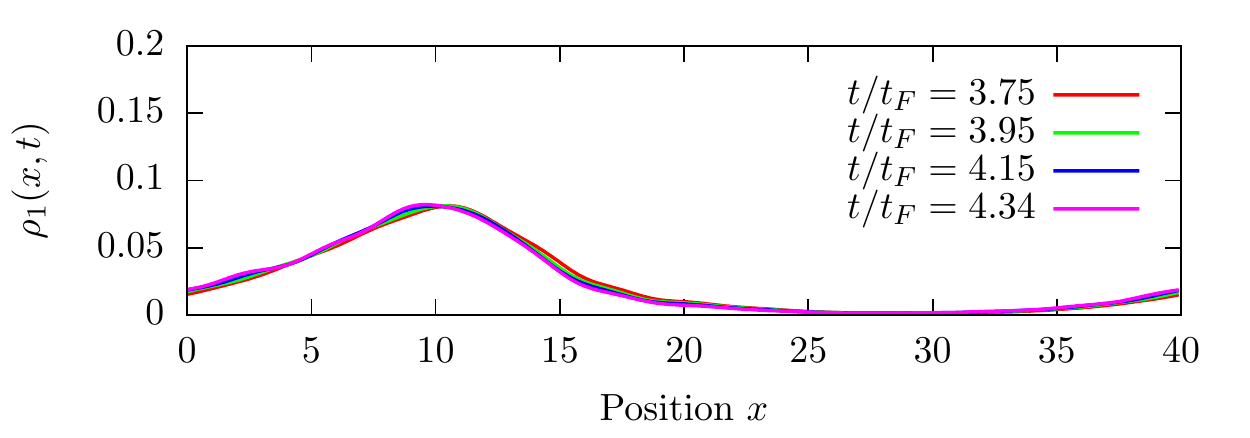} & 
\includegraphics[width=0.33\textwidth]{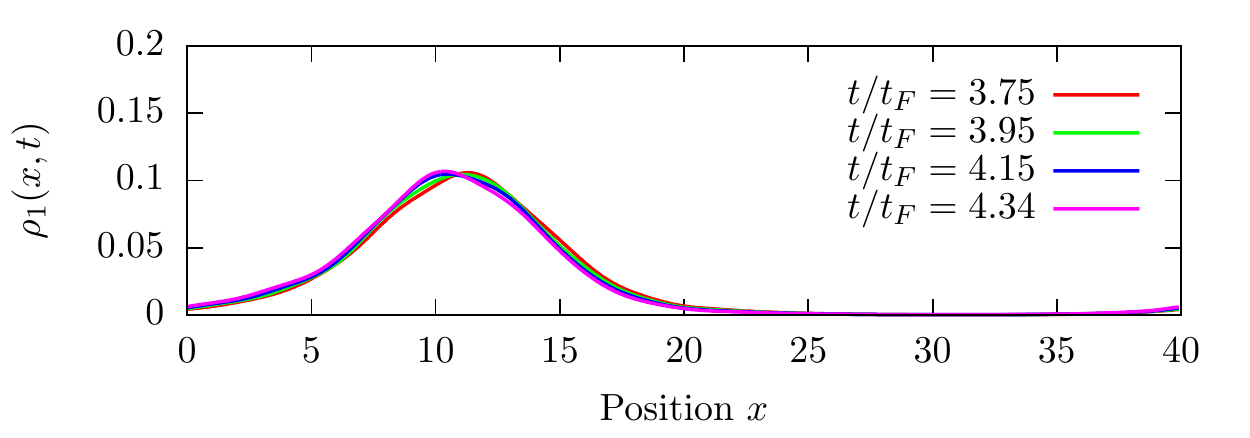} \\
\includegraphics[width=0.33\textwidth]{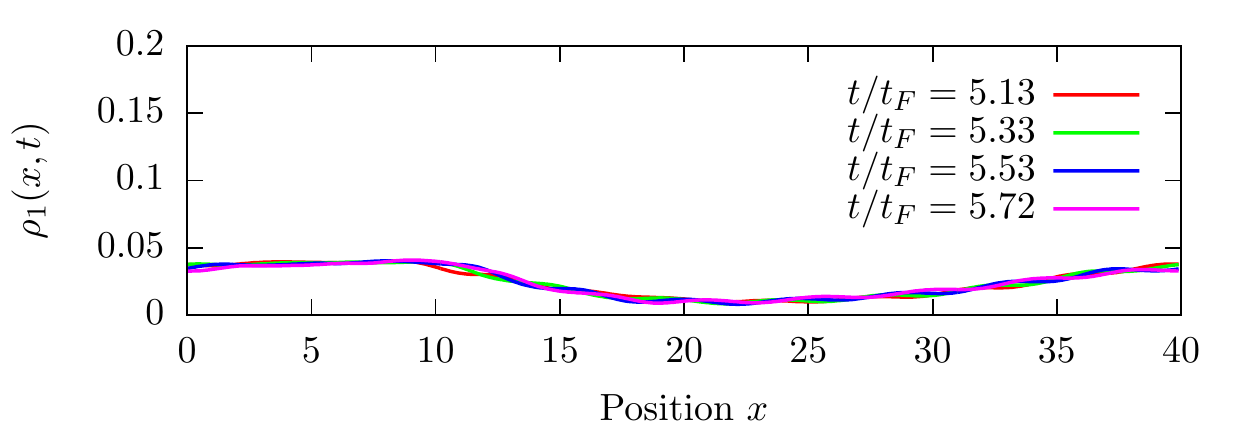} & 
\includegraphics[width=0.33\textwidth]{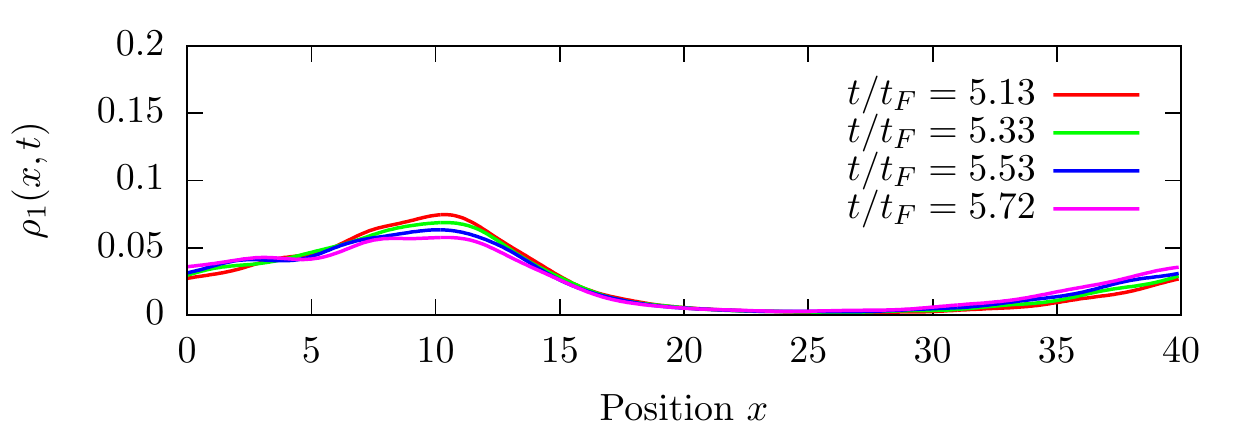} & 
\includegraphics[width=0.33\textwidth]{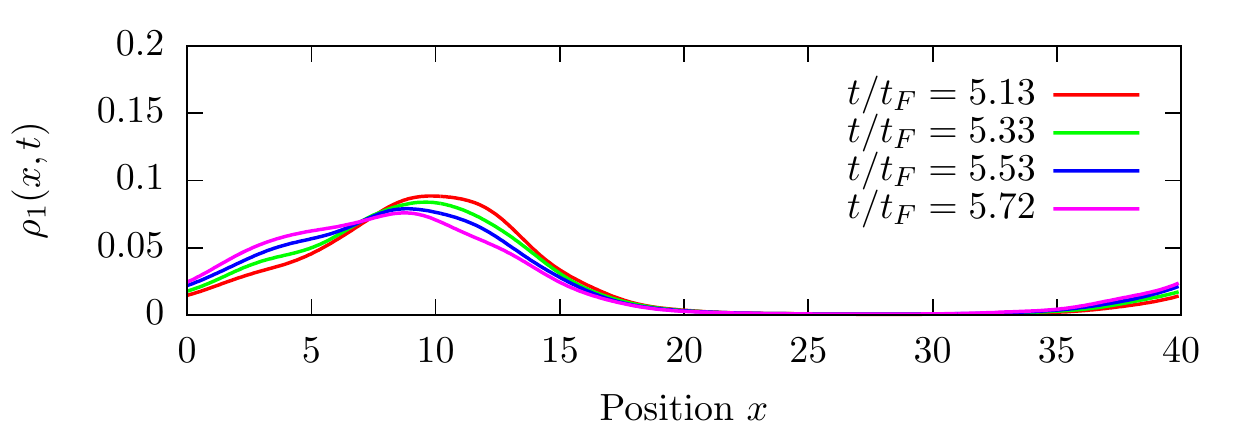}
\end{tabular}
\caption{Real time slices of the time-evolution of the impurity density for the three values of the interaction parameter $c=5,10,20$ from the initial state with $a_0^2 = 1.125$ and $Q = \pi$.}
\label{Fig:Slices}
\end{figure*}

Here we present additional data for the time-evolution of the initial state with $Q=0$ and $Q=\pi$. In Fig.~\ref{Fig:Q0}(a) we present the time-evolution for up to time $t\sim 20t_F$. As we saw in the main text, for times $2t_F \lesssim t \lesssim 7t_F$ the impurity undergoes arrested expansion: it is approximately stationary, with only small amplitude breathing oscillations. Following the arrested expansion, there is a period of rapid expansion, followed by a shorter quasistationary period and then subsequent expansion. In Fig.~\ref{Fig:Q0}(b) we present constant time cuts for short times $t \le 9 $ ($t \lesssim 3.5 t_F$) , which show the initial period of rapid expansion and subsequent arrested expansion. Figure~\ref{Fig:Slices} presents similar time cuts for the initial state with $Q=\pi$ for $c=5,10,20$. 

\subsection{Quantum Newton's cradle on the ring}

\subsubsection{Motion of the center of mass}

\begin{figure*}[ht]
\begin{tabular}{ll}
(a) & (b) \\ 
\includegraphics[width=0.45\textwidth]{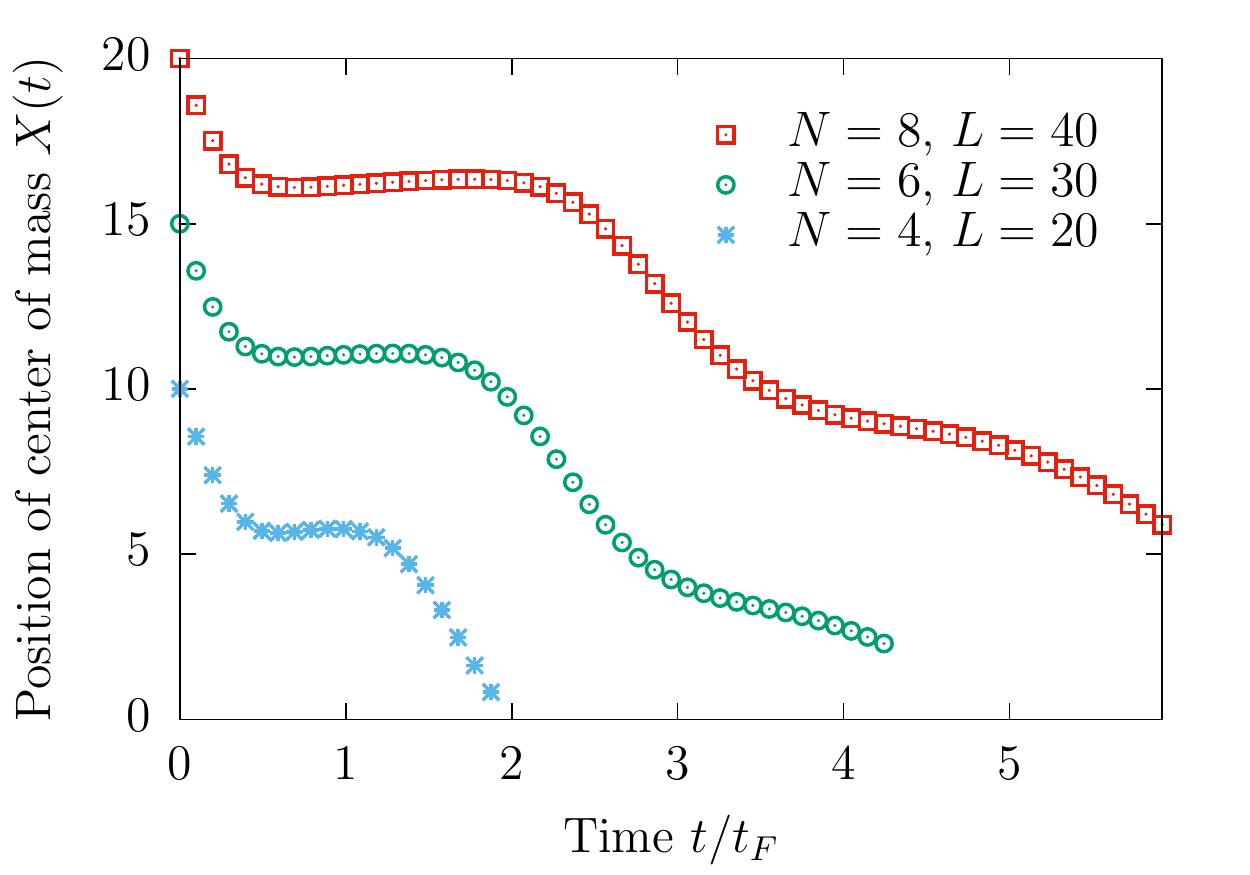} &
\includegraphics[width=0.45\textwidth]{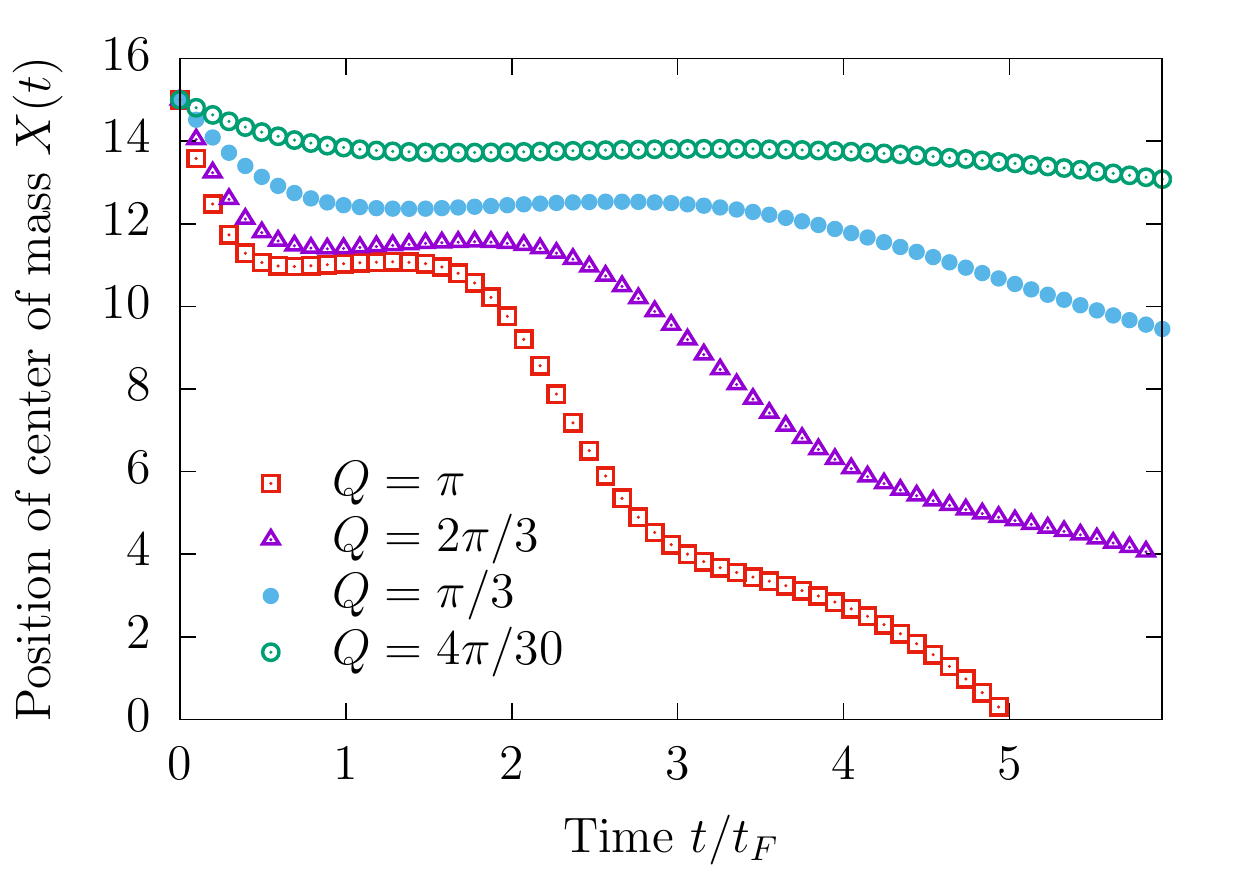}
\end{tabular}
\caption{Motion of the centre of mass coordinate $X(t)$ [as defined in Eq.~(8)] for the initial state $|\Psi(Q)\ra$ [defined in Eq.~(4)] with $x_0 = L/2$ and $Q=\pi$ for (a) a number of system sizes $L$ with fixed total particle density $N/L$ and interaction strength $c=10$; (b) a number of different initial center of mass momenta $Q$ for $L=30$, $N=6$ and $c=10$.}
\label{Fig:CoM_Ldep}
\end{figure*}

Here we provide support for our assertion that the motion of the center of mass coordinate can be explained in terms of a ``quantum Newton's cradle'' on the ring. The motion of the center of mass coordinate is presented in Fig.~3 of the manuscript, where we observe periods of rapid motion separated by approximately stationary plateaux (see also Figs.~\ref{Fig:Slices}). We have the following explanation for the observed behavior: as the impurity moves through the background gas, it collides and excites the background gas, with the excitations predominantly moving in the same direction as the impurity. The impurity continues to collide with the background gas until it has imparted all (or most) of its center of mass momentum, and subsequently the center of mass coordinate is (approximately) stationary. The excitations in the background gas propagate around the ring, until they once again reach the impurity and collide with it, imparting momentum and causing the center of mass of the impurity to once more move. This process then repeats. In support of this picture, in Fig.~\ref{Fig:CoM_Ldep}(a) we present the center of mass motion in three systems with different sizes and fixed particle density (this means the speed of sound in the system should be similar, up to small finite size effects), and we see that the time for which the center of mass is stationary is linearly dependent on the system size $L$. In Fig.~\ref{Fig:CoM_Ldep}(b) we present the center of mass motion for four different initial momenta $Q$, and we see the length of the plateau is inversely proportional to $Q$ at large $Q$ (the velocity of an excitation with momentum $Q$ is $v = Q/m$). Both of these results are consistent with the presented picture, where the deviation from the stationary plateau is driven by finite momentum excitations propagating around the ring.

\subsubsection{Behavior with variation of the interaction strength $c$}

How the behavior of the ladder motion of the center of mass changes with the interaction strength reveals the competition between interactions in the system and spreading of the impurity. An intuitive picture to have in mind is that of a liquid, which becomes `stiffer' with increasing interaction strength. At first blush, such a picture may seem to be inconsistent with the presented center of mass motion (see Fig.~3 of the main text), which appears to sharpen with weakened interactions. The observed behavior can be explained as follows. When the interaction strength $c$ is weak, the impurity spreads quickly and has almost completely delocalized by the second plateau. As a consequence, the second plateau is relatively flat and the transient region between the plateaux is broad. With strengthening interactions, the spreading of the impurity is hindered and the spreading has yet to finish by the time the second plateau is reached. The second plateau appears less stable for strong interactions as the impurity continues to slowly spread whilst approximately stationary, shifting the center of mass slightly. We give evidence for this picture in the remainder of this section. 

The behavior of the center of mass motion can be seen in Fig.~3 of the main text for $c=5,10,20$ (see also Fig.~\ref{Fig:CoM_Ldep} for $c=10$). We will first address the behavior of the transient regions and then the stability of the plateaux upon varying the interaction strength, showing that it is consistent with the intuitive picture of interactions creating a stiffer background gas. In the first transient region $t\lesssim t_F/2$ the  momentum of the impurity is imparted to the fluid: this happens more quickly for stiffer ($c$ larger) fluids. In the second transient region, the impurity is accelerated by collisions with the excitations of the background gas. The impurity immersed in the stiffer fluid is accelerated over a shorter period of time, and imparts its momentum back to the gas quicker, consistent with the intuitive picture. As a consequence, the transient region reduces in temporal extent with increasing interaction strength. 

\begin{figure*}
\begin{minipage}[b]{0.47\textwidth}
	\begin{flushleft} 
	(a) 
	\end{flushleft}
	\vspace{-4mm}
	\includegraphics[width=\textwidth]{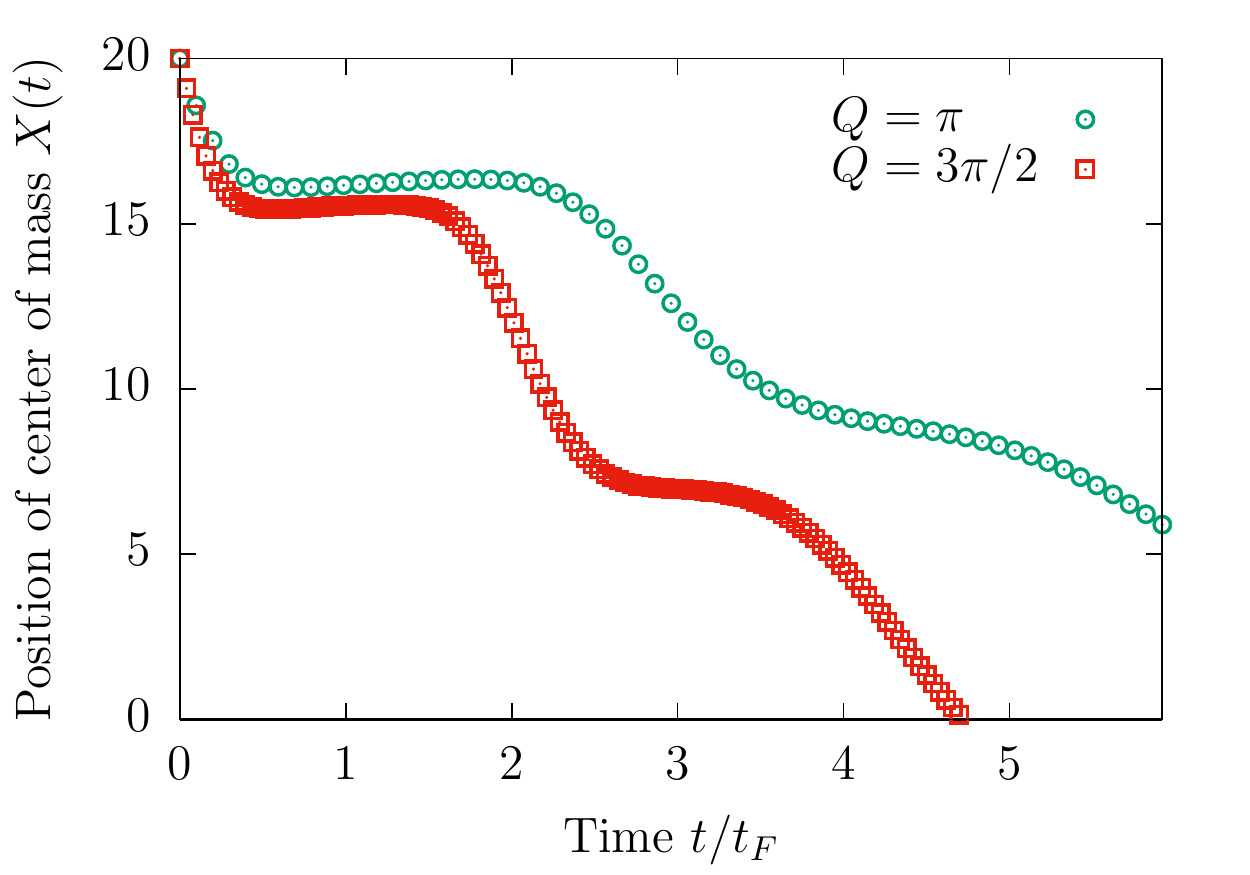} 
\end{minipage}
\hfill
\begin{minipage}[b]{0.48\textwidth}
	\begin{flushleft} 
	(b) 
	\end{flushleft}
	\vspace{-6mm}\includegraphics[width =\textwidth]{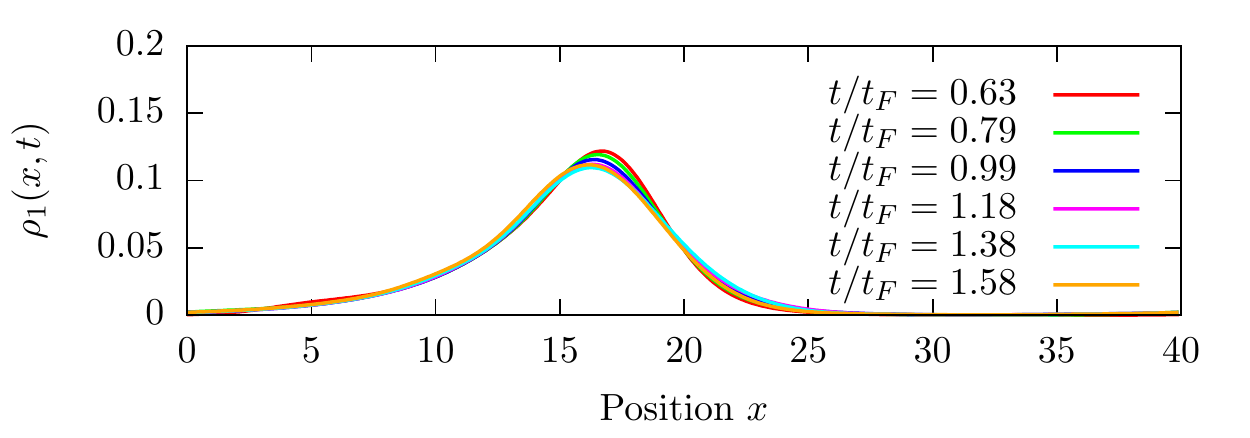} \\ \vspace{-6mm}
	\begin{flushleft} 
	(c) 
	\end{flushleft} 
	\vspace{-6mm}\includegraphics[width=\textwidth]{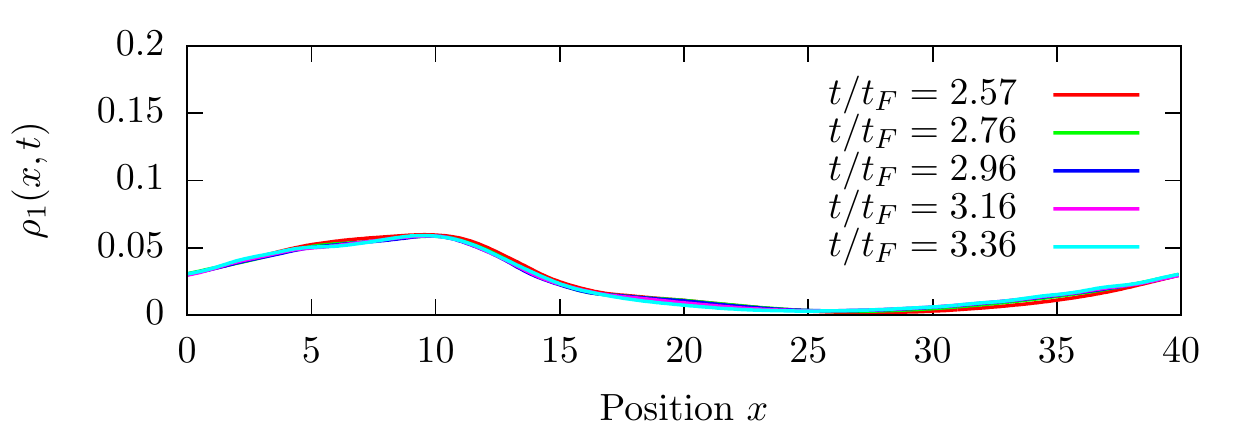} 
\end{minipage} 
\caption{(a) Motion of the center of mass of an impurity with $a_0^2 = 1.125$ and $Q = \pi,\ 3\pi/2$ for $N=8$ particles on the length $L=40$ ring with $c=10$. (b) The impurity density on the (b) first and (c) second plateau for $Q=3\pi/2$ ($cf$. the second column of Fig.~\ref{Fig:Slices} for $Q=\pi$).} 
\label{Fig:COMQ}
\end{figure*}

\begin{figure}
\includegraphics[width=0.45\textwidth]{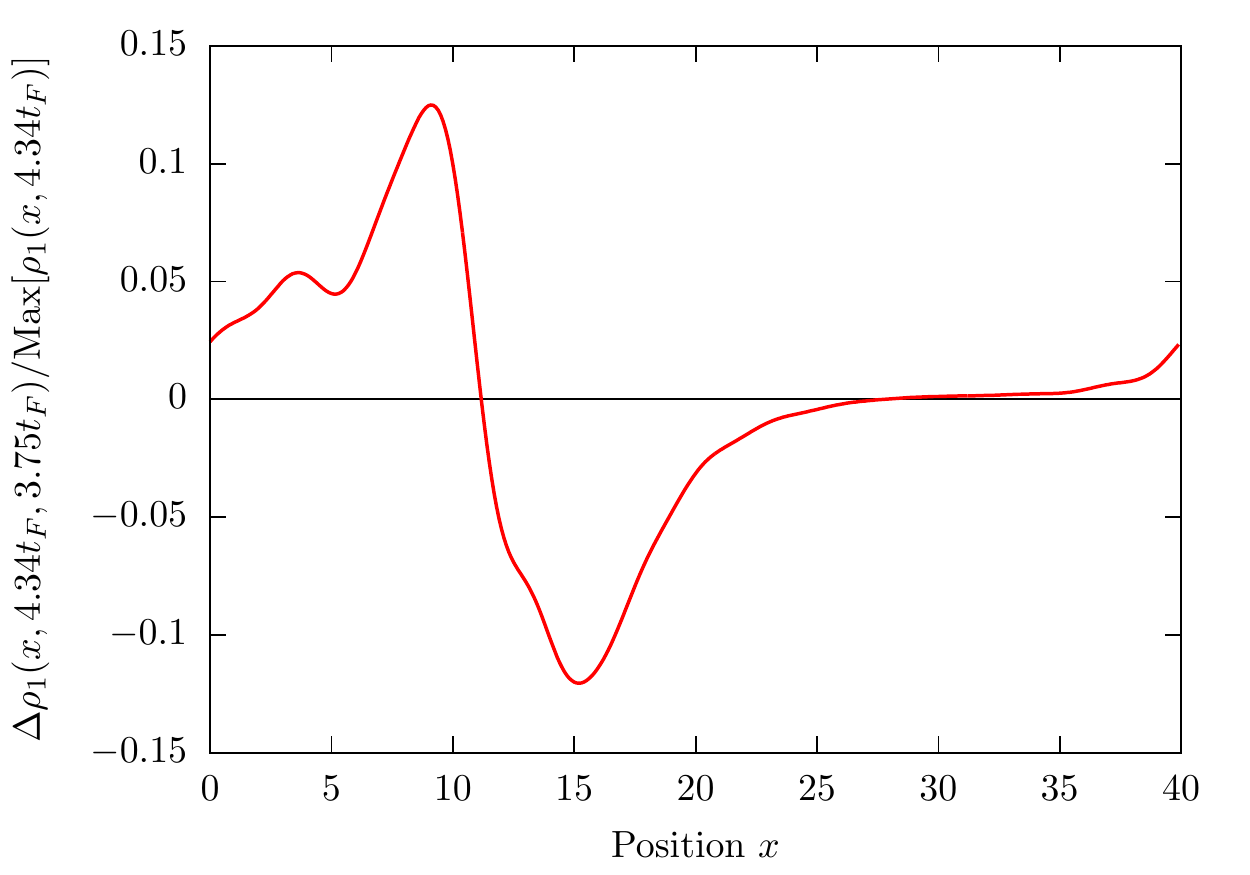}
\caption{The difference $\Delta\rho_1(x,t_1,t_2) = \rho_1(x,t_1) - \rho_1(x,t_2)$ in the density at times $t/t_F = 4.34$ and $t/t_F = 3.75$ for $c=20$, see Fig.~\ref{Fig:Slices}. We normalize to the maximum value of the density at time $t/t_F = 4.34$. A clear transfer of weight, from the right hand side of the impurity wave packet to the left with increasing time is seen, resulting in the drifting of the center of mass plateau shown in Fig.~3 of the main text.}
\label{Fig:Weight}
\end{figure}

Next, we consider the behavior of the second plateaux. It is useful to consider Fig.~\ref{Fig:Slices}; we see that the spreading of the impurity is suppressed with increasing interaction strength (this is particularly apparent in the second and fourth rows). We also see that the impurity has almost completely delocalized around the ring when $c=5$. This is important, as the local fluctuations in the density are proportional to the local derivative of the density, and hence fluctuations are suppressed with increasing delocalization of the impurity, which improves the stability of the plateaux. Increasing the momentum of the impurity also increases the rate at which the impurity delocalizes, and the plateaux are flatter, see Fig.~\ref{Fig:COMQ} [see also Fig.~\ref{Fig:CoM_Ldep}(b)]. Furthermore, Fig.~\ref{Fig:Slices} also shows that the slow drifting of the second $c=10,20$ plateau corresponds to small changes in the shape of the impurity, corresponding to a transfer of weight leftwards, see Fig.~\ref{Fig:Weight}. This transfer of weight, due to the slight asymmetric spreading of the impurity, causes the drifting of the second plateau. 

Delocalization of the impurity does not only increase the stability of the plateaux, it also leads to a smearing of the transient region, see Fig.~3 of the main text and Fig.~\ref{Fig:COMQ}(a). This is also consistent with our intuitive picture: excitations in the background gas now scatter on a increasingly extended object. Some excitations pass through the impurity, some scatter on the right or left of the impurity; the transient motion becomes more unclear and the transition between plateaux broadens [e.g., the peak velocity of the COM is reduced, see the inset of Fig.~3 of the main text and Fig.~\ref{Fig:COMQ}(a)].

\subsection{The diagonal ensemble}

\subsubsection{The impurity density in the diagonal ensemble}
\begin{figure}[b]
\includegraphics[width=0.45\textwidth]{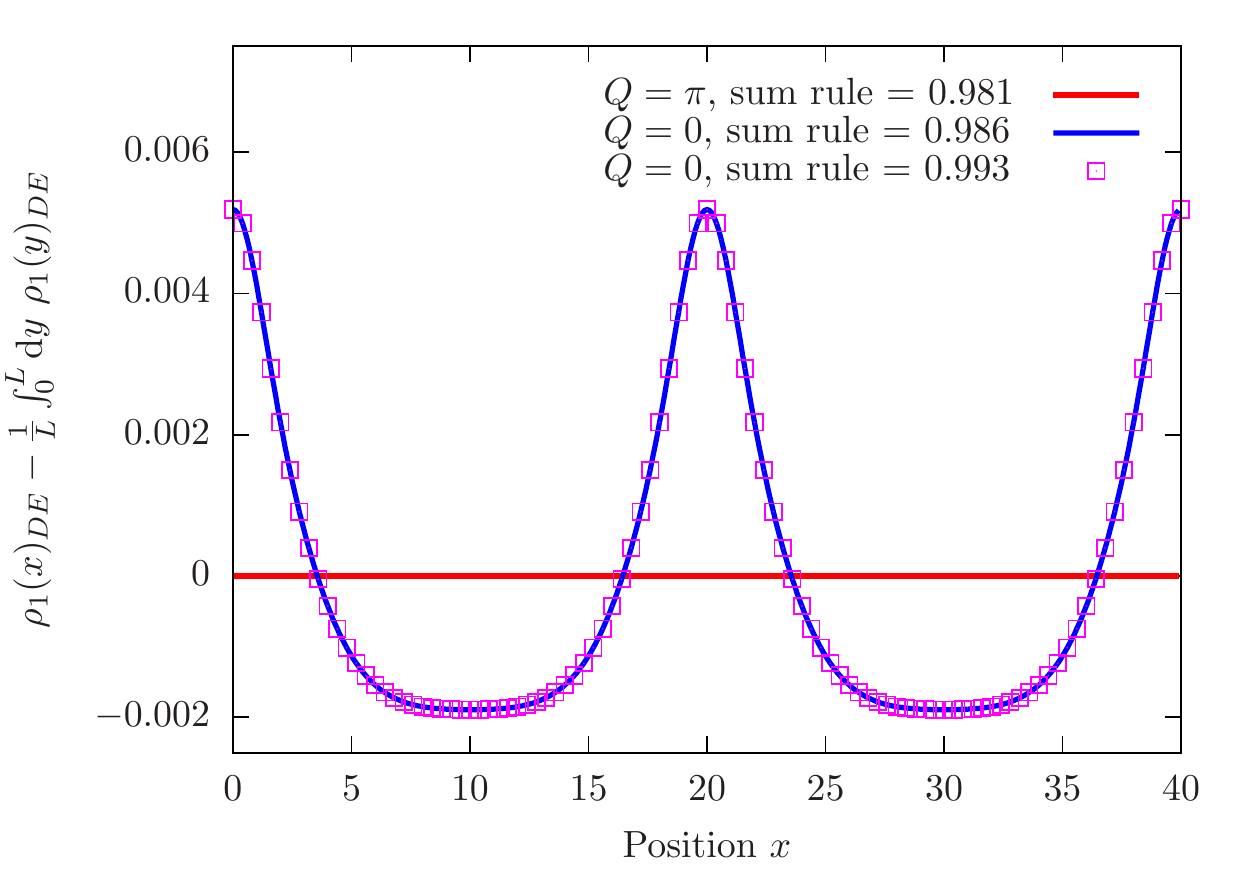}
\caption{Results for the impurity density in the diagonal ensemble~\fr{Eq:Diagonal_Ensemble} for the initial state~$|\Psi(Q)\ra$ [defined in Eq.~(4)] with momentum $Q=0,\pi$ and $a_0^2 = 1.125$ for $N=8$ particles on the length $L=40$ ring with interaction parameter $c=10$. The largest sum rule saturation requires 69532 states in each sum of Eq.~\fr{Eq:Diagonal_Ensemble}. Positions of the peaks are at $x_0$ and $x_0 + L/2$.}
 \vspace{-0.25cm}
\label{Fig:DE}
\end{figure}

To ascertain whether the impurity density becomes translationally invariant at long times after the impurity is injected, we compute the density profile in the diagonal ensemble
\bea
\rho_1(x)_{DE} = &&\sum_{\{k\};\mu} \sum_{\{p\};\lambda}  \delta_{E_k,E_p} e^{i(K_{p}-K_{k})x} 
\la\Psi(Q)|\{p\};\lambda\ra \nn
&& \times \la \{p\};\lambda| \Psi\dg_1(0)\Psi^{\phantom\dagger}_1(0) |\{k\};\mu\ra \la\{k\};\mu|\Psi(Q)\ra  ,\nn
\label{Eq:Diagonal_Ensemble}
\eea
which follows from a stationary phase argument in the long time limit. Herein we assume the diagonal ensemble coincides with the long time limit. Representative results for $N=8$ particles on the length $L=40$ ring with interaction parameter $c=10$ are shown in Fig.~\ref{Fig:DE}. There, we see that the diagonal ensemble result for the density with initial $Q=0$ is not translationally invariant, whilst for $Q=\pi$ the density profile appears much closer to constant. This provides strong evidence that for sufficiently large $Q$ the impurity is almost completely delocalized around the ring in the long-time limit.

\begin{figure}[b]
\begin{tabular}{l}
(a) \\ 
\includegraphics[width=0.45\textwidth]{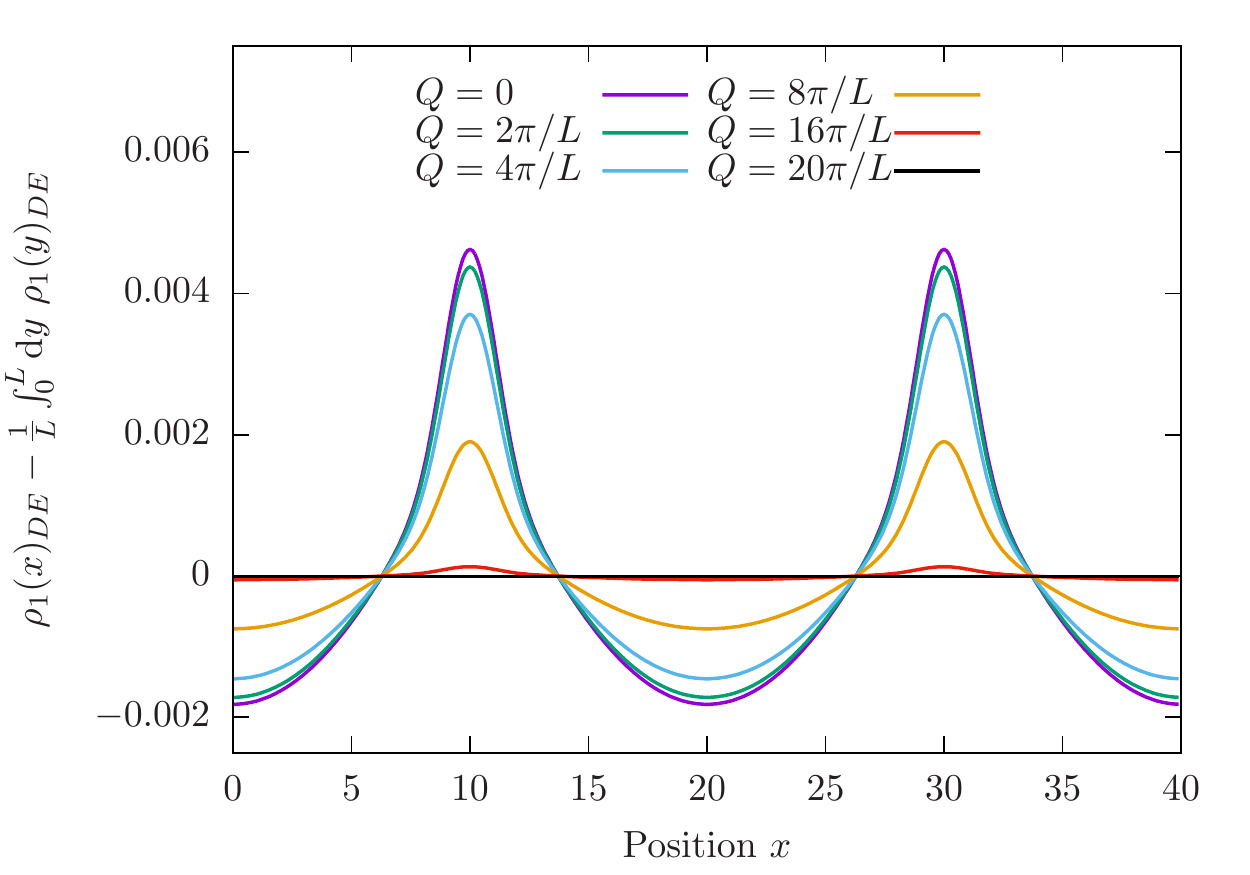} \\
(b) \\
\includegraphics[width=0.45\textwidth]{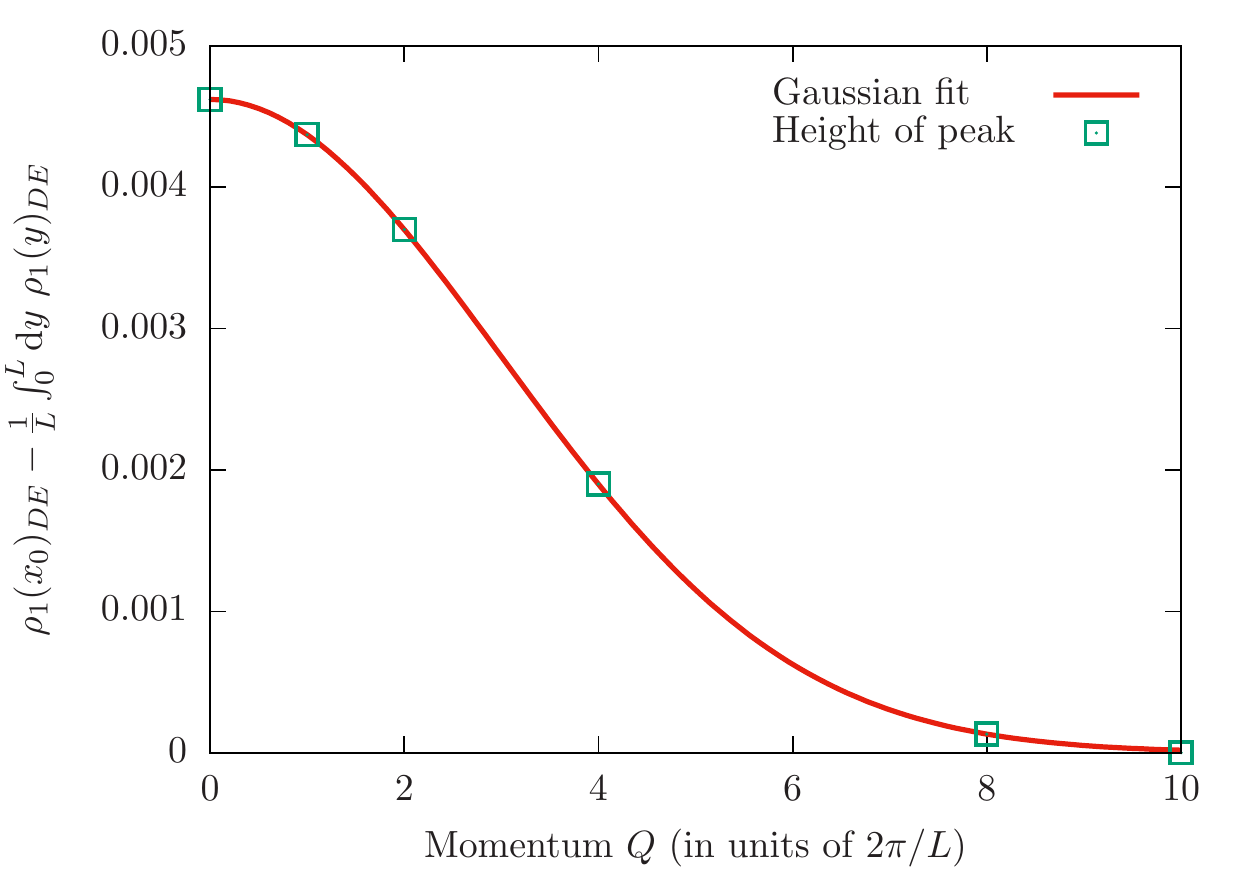} 
\end{tabular}
\caption{(a) Deviation of expectation value of the impurity density operator in the diagonal ensemble~\fr{Eq:Diagonal_Ensemble}  from the translational invariant case for the initial state with momentum $Q$ and $x_0 = L/4$. Results are for $N=4$ particles on the ring of length $L=40$ with interaction parameter $c=10$; sum rules are saturated to at least $0.997$ and we see now change in the extent of translational symmetry breaking with increasing saturation of the sum rule. (b) Maximum value of the deviation at $x=x_0$ as a function of $Q$ with a Gaussian fit. Note that the peaks in (a) are shifted compared to Fig.~\ref{Fig:DE} solely due the change in $x_0$.}
\label{Fig:DE_Q}
\end{figure}

One important question raised by Fig.~\ref{Fig:DE} is whether there is a sharp transition or a smooth reduction in the extent to which translational symmetry is broken with increasing $Q$. In Fig.~\ref{Fig:DE_Q}(a) we present the impurity density in the diagonal ensemble for a number of initial momenta $Q$ and $N=4$ particles, and we show that the severity of the translational symmetry  breaking is smoothly reduced as a Gaussian in the momentum of the initial state in Fig.~\ref{Fig:DE_Q}(b). Formally, this means that translational invariance is \textit{only recovered in the $Q\to\infty$ limit}. However, in a practical sense, translational symmetry is restored for sufficiently large $Q$ for a finite precision measurement. 

The origin of this Gaussian scaling can easily be explained by a degeneracy in our system: for each Bethe state $|\{k_1,k_2,\ldots,k_N\};\mu\ra$ there exists a state with the same energy and opposite momentum $|\{-k_1,-k_2,\ldots,-k_N\};-\mu\ra$. The diagonal ensemble~\fr{Eq:Diagonal_Ensemble} now contains two types of terms: diagonal matrix elements which sum to $1/L$ and off-diagonal terms $\delta\rho_1(x)_{DE}$ which break  translational invariance: 
\bea
\delta\rho_1(x)_{DE} &=& \sum_{\{k\};\mu} \la \Psi(Q)|\{k\};\mu\ra \la \{-k\};-\mu|\Psi(Q)\ra e^{2iK_{k}x} \nn
&& \times  \la \{k\};\mu| \Psi\dg_1(0)\Psi^{\phantom\dagger}_1(0) |\{-k\};-\mu\ra + \ldots 
\label{Eq:DE_breaking}
\eea
Here the ellipses denote other terms arising from other (possible) degeneracies. The overlap between the initial state and the Bethe states $\la\Psi(Q)|\{k\};\mu\ra$ is weighted by a Gaussian factor $\propto\exp(-Q^2)$. This Gaussian term will be present in \textit{any terms} which break translational invariance, and hence the extent to which translational symmetry is broken is smoothly suppressed with increasing $Q$, as observed in Fig.~\ref{Fig:DE_Q}(b). 

When considering the center of mass coordinate motion, we do not see a qualitative difference for cases in which the diagonal ensemble result is (almost) translationally invariant and those in which the translational invariance is more strongly broken, see Fig.~\ref{Fig:CoM_Ldep}(b).

\subsubsection{The momentum of the impurity in the diagonal ensemble}
Having computed the diagonal ensemble result for the density of the impurity, we now turn our attention to computation of the momentum of the impurity. We define the momentum of the impurity as
\be
K(t) = \sum_p\ p\ \la \Psi(Q)|e^{iHt} \Psi\dg_{1,p} \Psi_{1,p} e^{-iHt}|\Psi(Q)\ra,
\ee
where $\Psi_{1,p} = \frac{1}{L} \int \rd x e^{-ipx} \Psi\dg_1(x)$ is the momentum space annihilation operator for a boson of species $1$. In the $t\to\infty$ limit we assume that this is given by the diagonal ensemble and as  $\Psi\dg_{1,p}\Psi_{1,p}$ conserves momentum, the diagonal ensemble result is
\bea
K_{DE} = \sum_{\{k\};\mu} \sum_p\ && p\ \la \Psi(Q)|\{k\};\mu\ra\la \{k\};\mu|\Psi(Q)\ra \nn
&& \times \la\{k\};\mu| \Psi\dg_{1,p}\Psi_{1,p} | \{k\};\mu\ra. 
\eea
Fourier transforming to real space operators, and inserting the resolution of identity over one-component Lieb-Liniger eigenstates, we find
\bea
K_{DE} = &&\sum_{\{k\};\mu} \sum_{\{q\}} \Big( K_{k} - K_{q} \Big)\nn
&&\times  \Big| \la \Psi(Q)|\{k\};\mu\ra \la \{k\};\mu| \Psi\dg(0)|\{q\}\ra\Big|^2, 
\eea
which is expressed in terms of known matrix elements. Using the previously discussed symmetry of the Bethe states (and a similar property for the one-component states) and the properties of the matrix elements of the creation operators, we can write the momentum of the impurity in the diagonal ensemble as 
\bea
K_{DE} \propto &&\sum_{\{k\};\mu}\sum_{\{q\}} \Big( K_{k} - K_{q} \Big)\Big(e^{-a_0^2(Q+K_k)^2} - e^{-a_0^2 (Q-K_k)^2}\Big)\nn 
&&\times\Big| \la \Omega|\Psi(0)|\{k\};\mu\ra\la \{k\};\mu|\Psi\dg(0)|\{q\}\ra\Big|^2
\eea
for the one-component ground state $|\Omega\ra$ containing an odd number of particles. We immediately see that for $Q=0$ the momentum of the impurity remains at zero at all times (consistent with the $Q=0$ time-evolution). 

\newpage

\end{document}